\documentclass[a4paper,usenatbib,useAMS]{mnras}

\usepackage[labelfont=Large]{caption}
\usepackage{subcaption}
\usepackage{graphicx}
\usepackage{setspace}
\usepackage{natbib}
\usepackage{color}
\usepackage{amsmath,amssymb}
\usepackage{times}

\title[Calibrating the BHB Star Distance Scale]{Calibrating the BHB Star Distance Scale and the Halo Kinematic Distance to the Galactic Centre}

\author[N. D. Utkin et al.]{
Nikita D. Utkin,$^{1,2,3}$\thanks{E-mail: nikitautkin@bk.ru}
Andrei K. Dambis,$^{1}$
\\
$^{1}$ Sternberg Astronomical Institute, M.V.Lomonosov Moscow State University, Universitetskii pr. 13, Moscow, 119991 Russia \\
$^{2}$ Faculty of Physics, Moscow State University, Leninskie Gory 1, bld. 2, Moscow, 119234 Russia \\
$^{3}$ Astro Space Center, Lebedev Physical Institute, Russian Academy of Sciences, 84/32 Profsoyuznaya st., Moscow, GSP-7, 117997, Russia}

\date{Accepted XXX. Received YYY; in original form ZZZ}

\pubyear{2020}

\begin{document}
\label{firstpage}
\pagerange{\pageref{firstpage}--\pageref{lastpage}}
\maketitle

\begin{abstract}
We report the first determination of the distance to the Galactic centre based on the kinematics of halo objects. 
We apply the statistical-parallax
technique to the sample of $\sim$~2500  Blue Horizontal Branch (BHB) stars compiled by \citet{XueBHB11} 
to simultaneously constrain the { correction factor to the photometric distances of BHB stars as reported 
by those authors} and the distance to the Galactic centre to find $R$~=~8.2~$\pm$0.6~kpc. 
We also find that the average velocity of our BHB star sample in the 
direction of Galactic rotation, 
$V_0$~=~-240~$\pm$~4~km/s, is greater by about 20~km/s in absolute
value than the corresponding velocity for halo RR Lyrae type stars ($V_0$~=~-222~$\pm$~4~km/s) in the Galactocentric
distance interval from 6 to 18~kpc, whereas the total ($\sigma V$) and radial ($\sigma r$) velocity dispersion 
of the of the BHB sample  are smaller by about 40--45~km/s than the corresponding parameters of the velocity 
dispersion ellipsoid of halo RR Lyrae type variables. The velocity
dispersion tensor of halo BHB stars proved to be markedly less anisotropic than the corresponding
tensor for RR Lyrae type variables: the corresponding anisotropy parameter values are equal to
$\beta_{BHB}$~=~0.51$\pm$~0.02 and  $\beta_{RR}$~=~0.71~$\pm$~0.03, respectively.

\end{abstract}




\section{Introduction}

The distance of the Sun to the nearest (our own) galaxy - i.e., the Galactic centre, $R_0$, is a fundamental scale factor determining such 
physical parameters of the Milky Way as  its mass and luminosity, as well as the mass distribution within it and hence the size and shape of orbits 
of various Galactic objects \citep{DGB16}. As the above authors point out, $R_0$ estimates fall into three main categories --- direct distance 
determinations, centroid-based determinations, and, finally, kinematic-based Galactic-centre distance determinations. The latter
mostly 
derive from the kinematics of Population-I (i.e., Galactic-disk) objects like Cepheids, open clusters, supergiant stars,  and masers 
\citep{Bobylev, ZhuShen, Reid14, BB14a, BB14b, Rastorguev17} as the distance to the kinematic centre of the velocity field incorporating
circular rotation and spiral-wave perturbations. Population-II kinematics-based Galactic centre distance determinations usually
involve the determination of the solar velocity with respect to some Population-II tracer sample and comparing it with some adopted
angular-velocity value at the solar Galactocentric distance \citep{K18}. 

The aim of this paper is to simultaneously determine both the distance from the Sun to
the kinematic centre of the halo velocity field and the distance-scale correction factor by applying the maximum-likelihood version of the statistical-parallax technique
to a sample of $\sim$~2500 purportedly clean 
Galactic-halo blue horizontal-branch (BHB) stars with full 6D data (sky positions, relative distances, proper motions, and radial velocities). 
This is the first
determination of the distance to the Galactic centre based on the assumption that velocity dispersion tensors at all halo points are aligned along the local
direction toward the Galactic centre and have the same shape and size. This determination has become possible owing to the unprecedented accuracy of proper motions provided by the second data release of 
Gaia astrometric space mission \citep{Gaia,DR2} and the fact that inside 25--30~kpc the velocity distribution of halo stars is highly lobe-shaped and radially 
anisotropic, being dominated by the so-called "Gaia Sausage" component found from the analysis of the kinematics of BHB stars and RR Lyrae
type variables \citep{sausage, Iorio} (if the velocity dispersion ellipsoid were spherical and had the same size
irrespectively of the Galactocentric distance it would hardly matter where to ``put'' the Galactic centre and this
parameter would remain practically impossible to constrain by the solution).

The layout of the paper is as follows. Section~\ref{sec:data} describes the data employed and the cuts applied to it. 
Section~\ref{sec:method}   briefly describes the method employed.  The next two sections describe the method
employed and the results obtained and, finally, Section~\ref{sec:conclusion} provides the  conclusions.

\section{Data}
\label{sec:data}

\begin{center}
\begin{figure*}
\includegraphics[width=\linewidth]{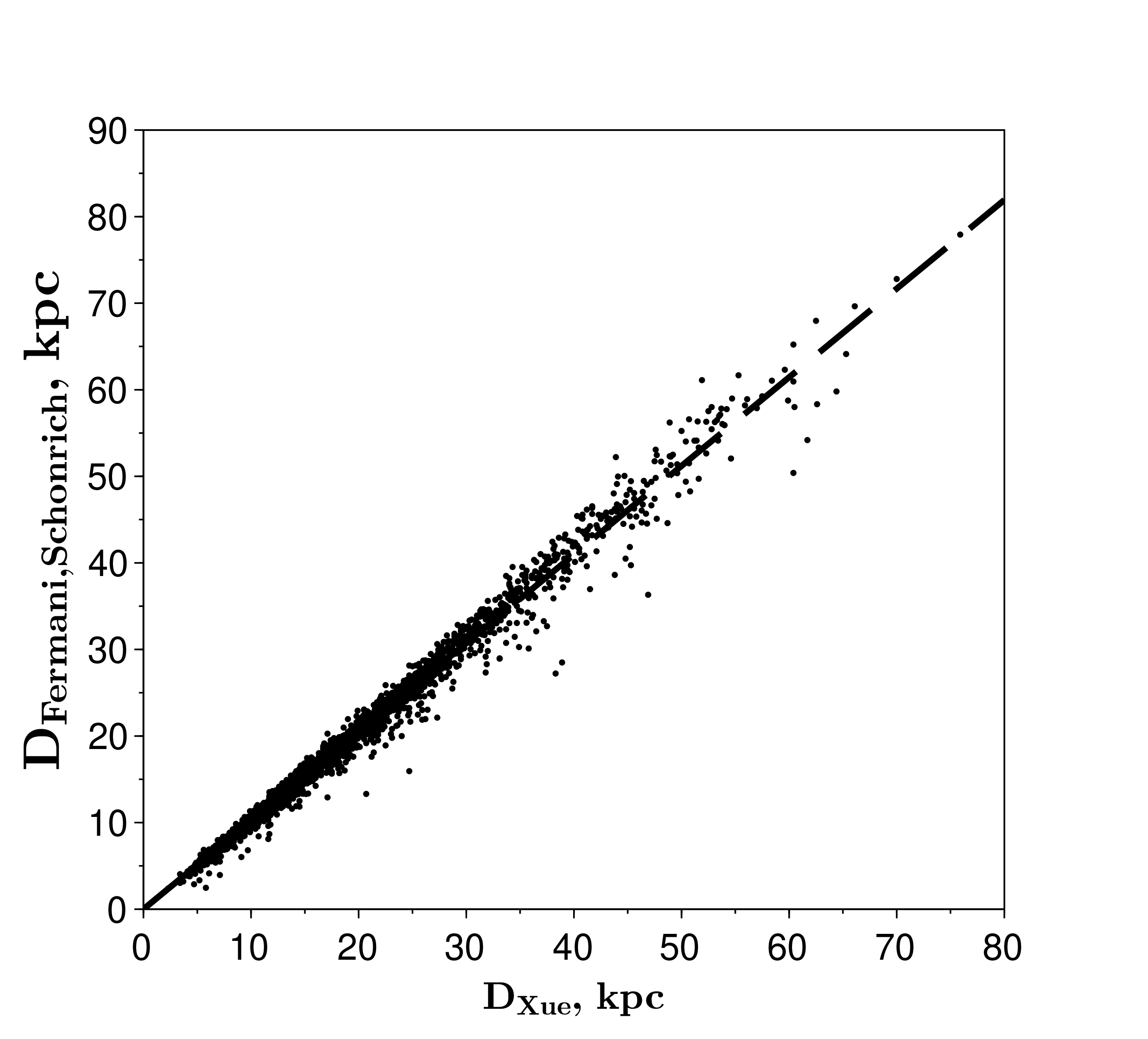}
\caption{ Comparison of the original BHB distances from \citet{XueBHB11} with the distances
computed using the calibration of \citet{Fermani2}. The straight line corresponds to the average scaling factor 
of $D_{Fermani,Schonrich}$/$D_{Xue}$~=~1.0288.  }
\label{compare_distances}
\end{figure*}
\end{center}

\begin{figure*}
    \centering
    \begin{subfigure}[t]{0.49\textwidth}
        \centering
        \caption{\Large} \label{g_mag}
        \includegraphics[width=\linewidth]{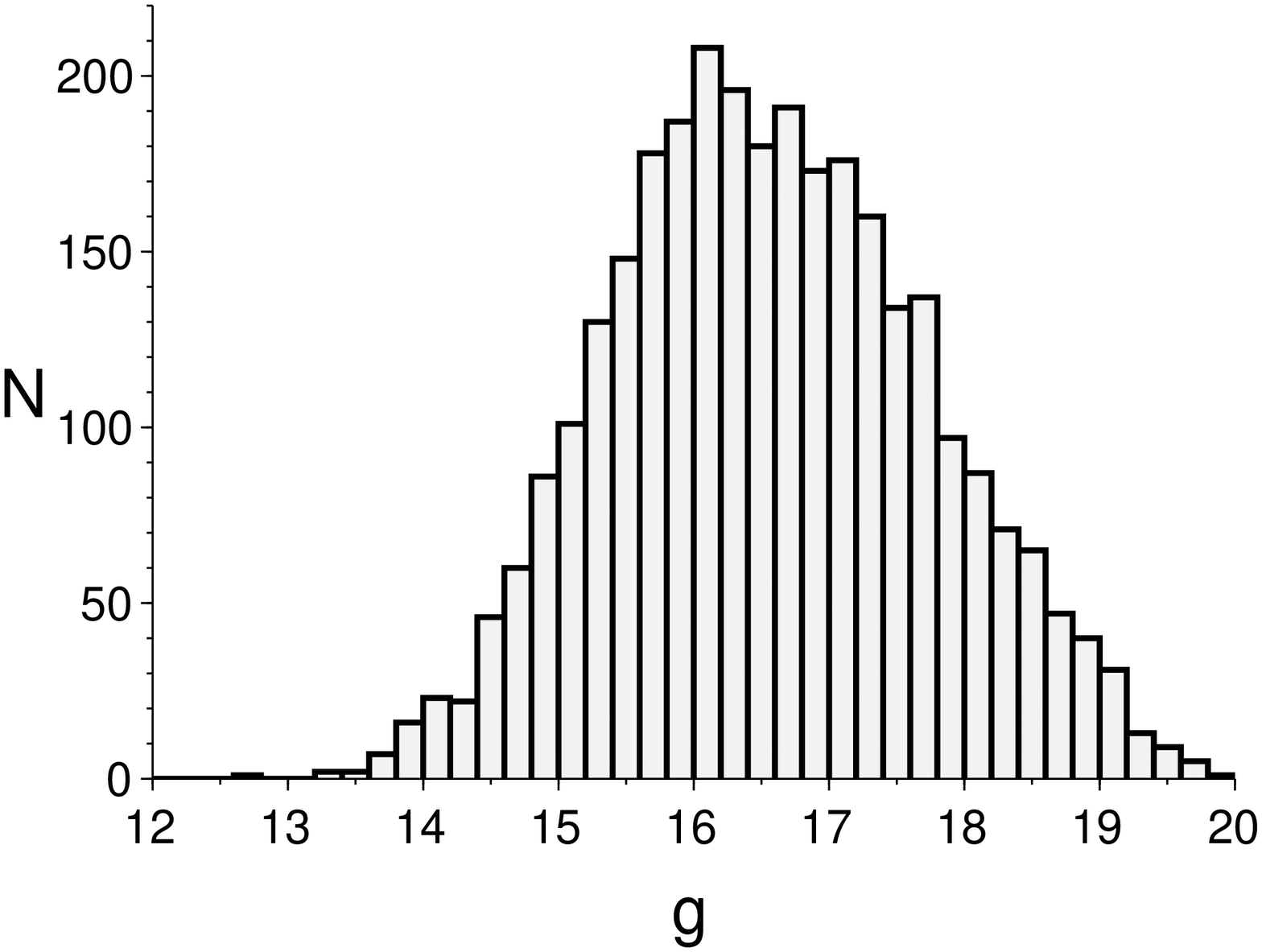} 
    \end{subfigure}
    \hfill
    \begin{subfigure}[t]{0.49\textwidth}
        \centering
        \caption{} \label{err_vr} 
        \includegraphics[width=\linewidth]{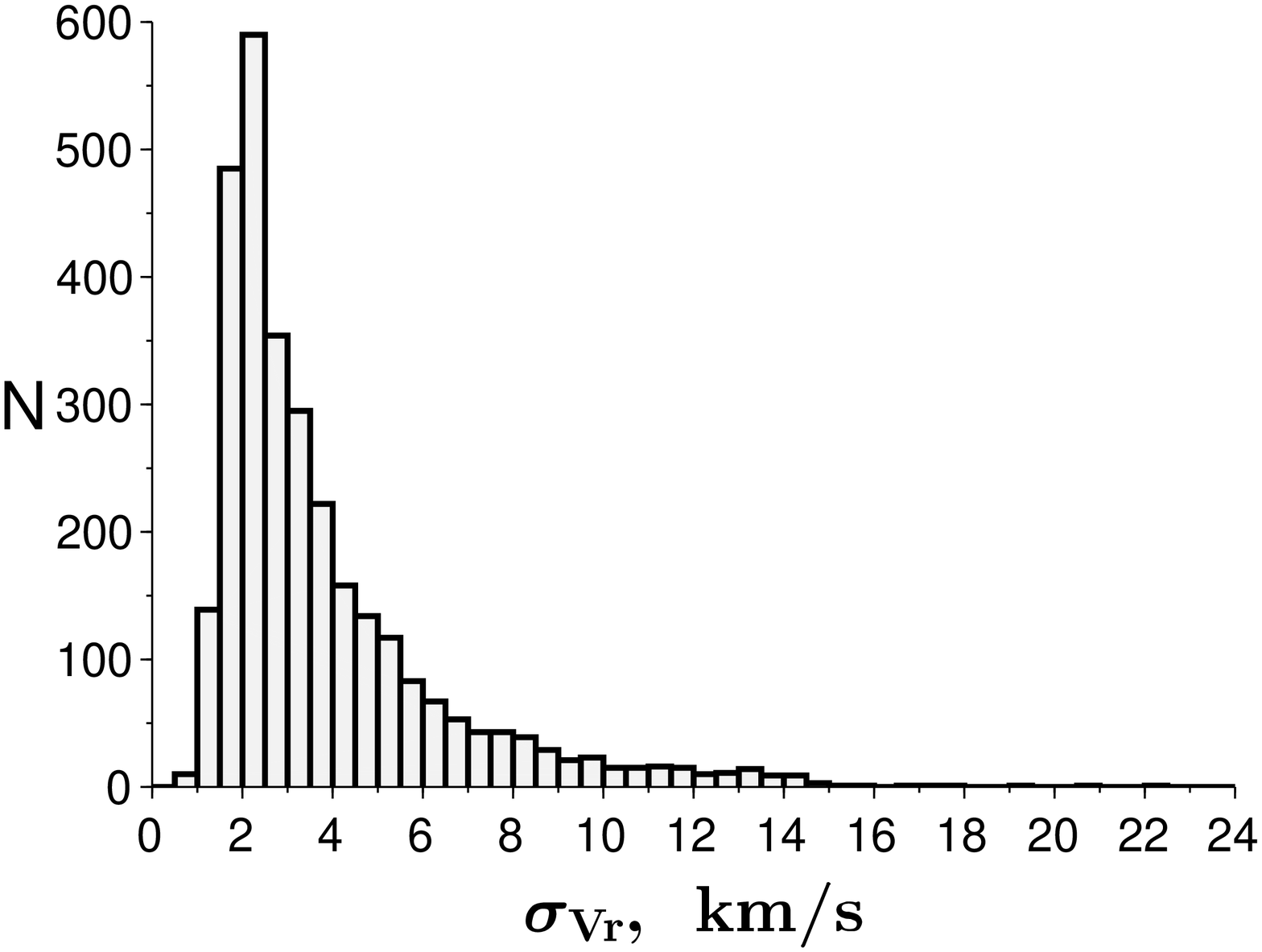} 
    \end{subfigure}

    \vspace{1cm}

\centering
    \begin{subfigure}[t]{0.49\textwidth}
        \centering
        \caption{} \label{err_pma} 
        \includegraphics[width=\linewidth]{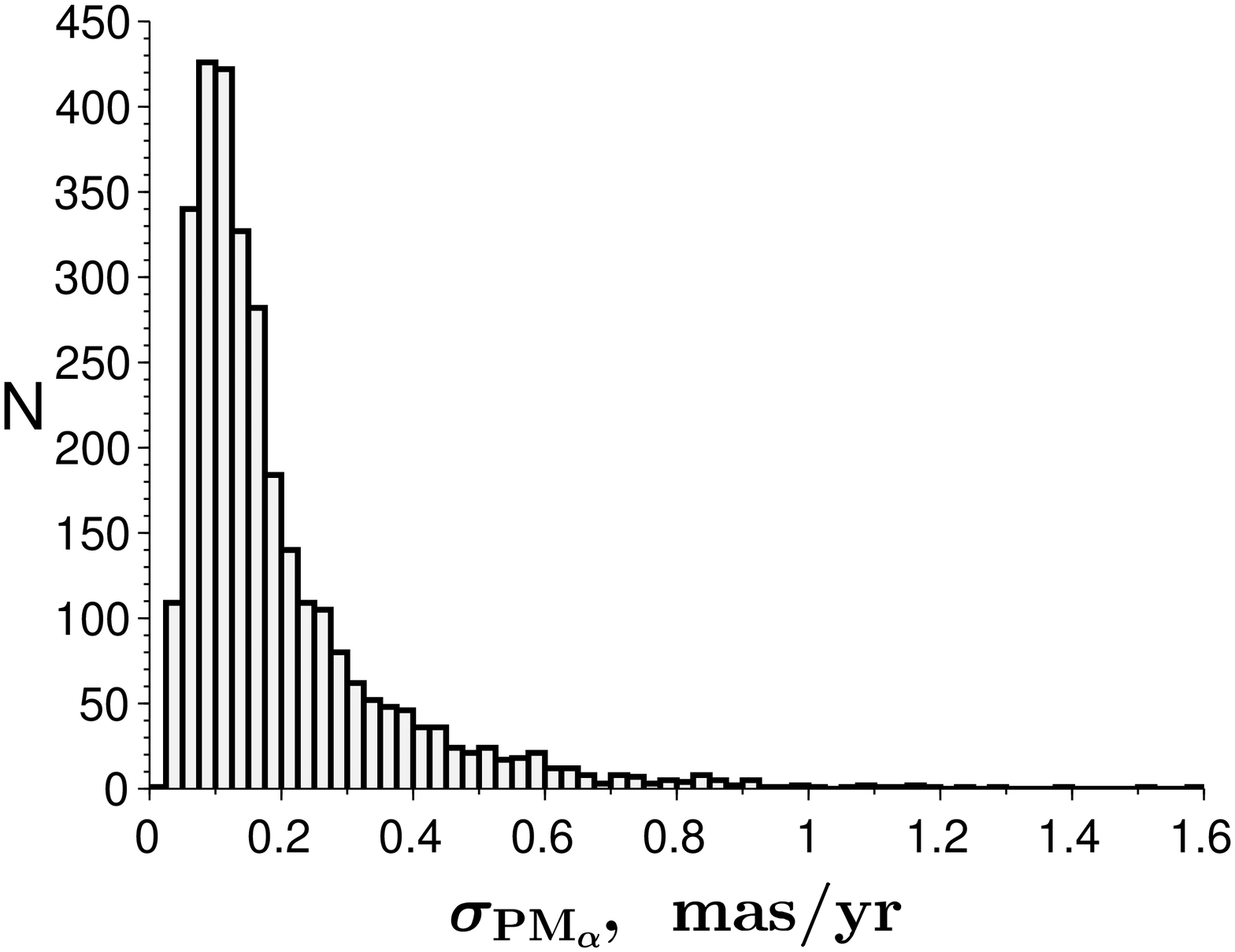} 
    \end{subfigure}
    \hfill
    \begin{subfigure}[t]{0.49\textwidth}
        \centering
        \caption{} \label{err_pmd} 
        \includegraphics[width=\linewidth]{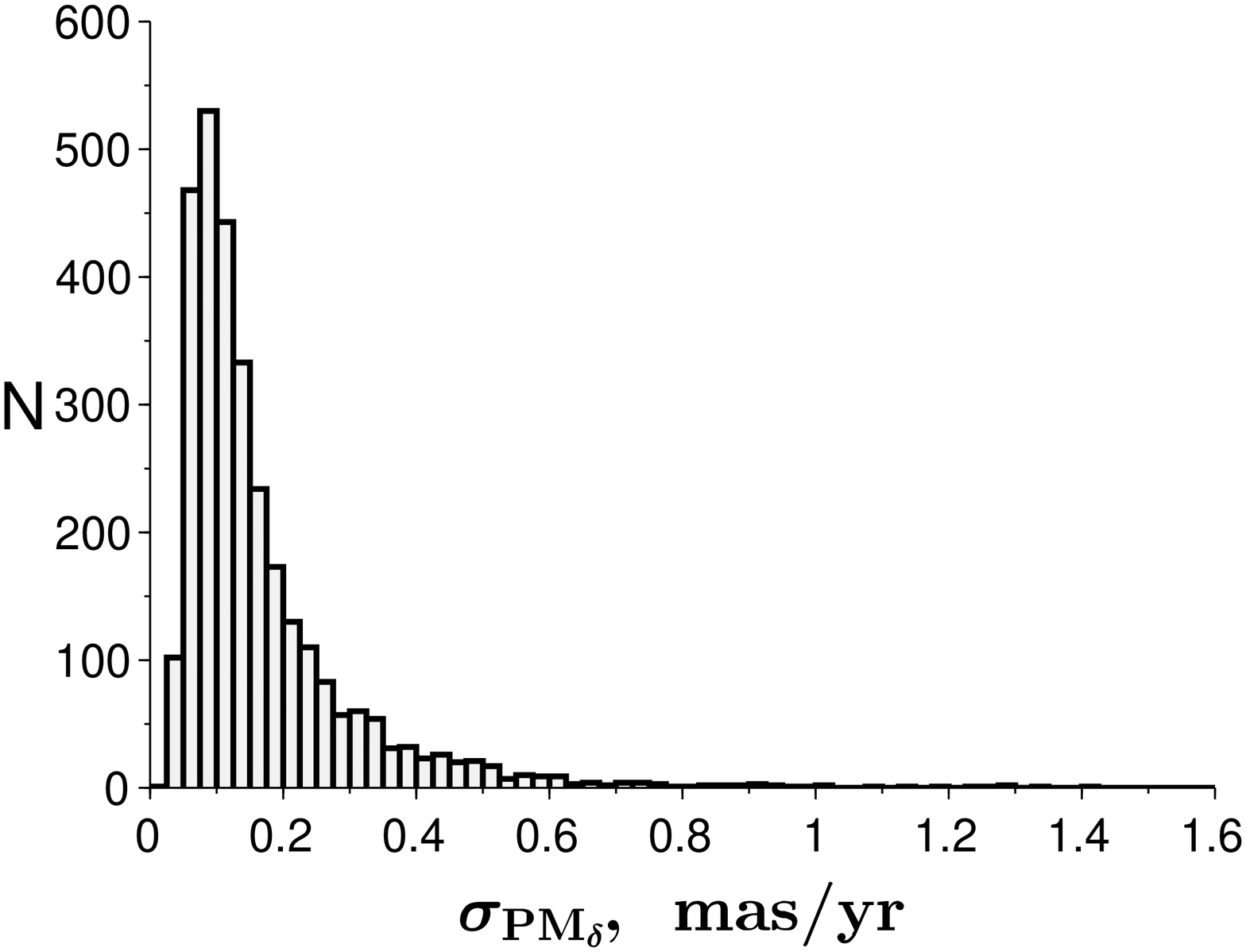} 
    \end{subfigure}

    \vspace{1cm}

    \begin{subfigure}[t]{\textwidth}
    \centering
        \caption{} \label{feh} 
        \includegraphics[width=\linewidth]{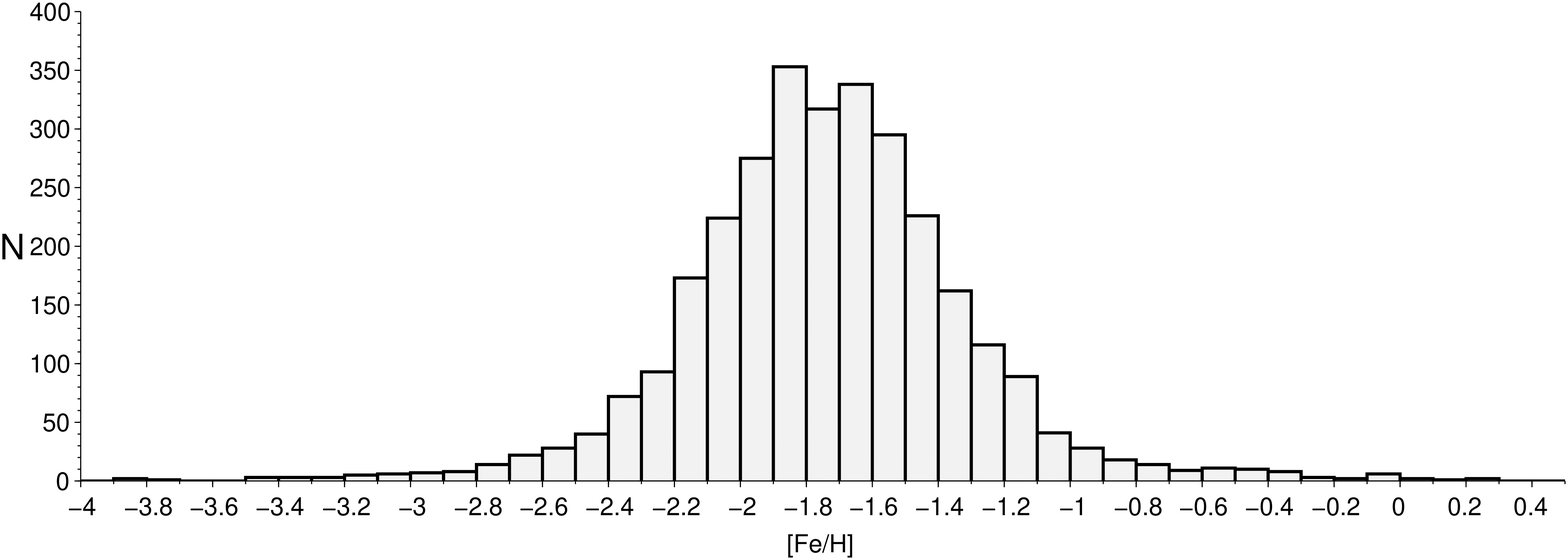} 
    \end{subfigure}
    \caption{ Diagnostic diagrams for the BHB sample employed: distribution of SDSS $g$-band magnitudes (a) and radial-velocity errors $\sigma_{Vr}$ (b);
              errors of the  proper-motion components $\sigma_{PM_{\alpha}}$ (c) and $\sigma_{PM_{\delta}}$  (d) in right ascension and declination, 
              respectively, and the distribution of metallicities [Fe/H] (e).}
\end{figure*}

Like \cite{sausage}, we use  the catalogue of 4,985 BHB stars compiled by \cite{XueBHB11} { based on SDSS DR8 \citep{DR8} data} as our initial sample of halo kinematic
tracers.  The spectra employed by \cite{XueBHB11} to identify BHB stars and determine their parameters were acquired within
the framework of SEGUE program, which was a subsurvey of SDSS-II project whose data were distributed as part of SDSS DR8. 
The radial velocities of the stars were determined via SEGUE Stellar Parameter Pipeline, which was used to process the 
calibrated spectra generated by the standard SDSS spectroscopic reduction pipeline \citep{Stoughton}.
{ Particular parameters of Balmer-line profiles needed to distinguish BHB stars from other stars of similar temperature
--- blue stragglers and main-sequence stars --- were computed by \cite{XueBHB11}  directly from SDSS spectra. These include two
parameters of the H$\delta$ line --- its width $D_{0.2}$  at 20\% below the local continuum and its flux $f_m$ relative to the continuum, 
and the parameters $b$ and $c$ of the S\'{e}rsic profile, $\rm y = 1.0 -
a\rm \exp{\left[-\left(\frac{|\lambda-\lambda_0|}{b}\right)^c\right]}$ of the H$\gamma$ line \citep{XueBHB08, Sirko04}.} 
{ We further supplement these data with} Gaia DR2 \citep{Gaia, DR2} proper motions to obtain an initial list of 4537 BHB stars with complete 6D phase-space
information. { Furthermore, to prevent eventual biases in the  the shape of the velocity ellipsoid, we
decontaminate our sample by removing stars that might belong to the well-known
Sagittarius (Sgr) stream. We do it by eliminating the objects that fall within its sky region as defined by \cite{Deason11} (see \cite{Fermani}).
We do not use a more elaborate approach involving the use of kinematic data for identifying stream members \citep{Antoja, Ibata}
because our statistical-parallax method operates with the likelihood function in the velocity space and tampering with the 
kinematic data may produce extra biases that we are just trying to avoid. The simple approach of masking the stream in the sky is
less likely to produce extra bias. We defer a more detailed  analysis with explicitly incorporating the Sagittarius stream 
into our kinematical model to a future study to be based on a more extensive tracer sample.
}
{ We further exclude all objects within 5~kpc from the Galactic midplane to prevent contamination by thick-disk stars with their markedly
different kinematics (much lower velocity  dispersion components and fast rotation \citep{Layden, Dambis09, Dambis13}). 
The 5~kpc cutoff should be sufficient to provide a clean halo sample 
given that the scaleheight of the thick disk in the Milky Way is of about
0.9~kpc \citep{Juric}. Finally, we exclude all stars with total Galactic rest-frame velocities higher than 600~km/s since they
should be either escaping from our Galaxy or have erroneous data. The final sample has a size of 2582 stars.}
In their catalogue \cite{XueBHB11} provide sky positions, distance estimates, radial velocities, and radial-velocity errors for all stars of the sample, but
give no individual distance errors. However, they point out that { the quoted relative distance estimates are typically
accurate to within 5\%,} and it is this error that we adopt for all stars in our subsequent kinematical analysis.  These distances, however, are computed
without taking into account the BHB absolute magnitude dependence on metallicity. To see how this dependence may affect our results, we
also computed a solution using the distances based on the absolute-magnitude calibration proposed by \cite{Fermani2} and expressed
as a function of $(g-r)_0$ and [Fe/H]. { To this end, we further
complete the data by adding the SPPP [Fe/H] estimates drawn from SDSS database \citep{Yanny}. Fig.~\ref{compare_distances} compares
the original distances from the catalogue of \cite{XueBHB11} with the distances based on the calibration of \cite{Fermani2}. The two distance sets 
can be seen to agree quite well once the scaling  factor is adjusted ($D_{Fermani,Schonrich}/D_{Xue}$~=~(1.0288~$\pm$~0.0008), with a scatter of
0.043). As we will see below, this is very close to the ratio  of the scaling factors delivered by the statistical-parallax method (1.024).
\cite{Fermani2} do not provide an estimate for individual errors of their  distances, but the discussion in that paper 
suggests that the fractional accuracy of their distances should be at least better than $\pm$~0.09 (9\%) corresponding to the absolute-magnitude
error of $\epsilon_{Mg}$~=~0.18. Fig.~\ref{g_mag} shows the distribution of
SDSS $g$-band magnitudes of the BHB stars of our sample. Fig.~~\ref{err_vr} shows the distribution of radial-velocity errors, $\sigma_{Vr}$,
Figs.~~\ref{err_pma} and ~\ref{err_pmd} show the distributions of the errors of the proper-motion components in right ascension ($\sigma_{PM_{\alpha}}$) and 
declination ($\sigma_{PM_{\delta}}$), respectively, and
Fig.~\ref{feh} shows the distribution of the metallicity values [Fe/H].}

\section{Determining the Velocity Field Parameters, the distance-scale correction factor, and the solar Galactocentric distance}
\label{sec:method}

{ Our tool of choice for inferring the kinematic properties and the distance-scale correction factor for the sample of stars is the method of statistical
parallax in its maximum-likelihood version suggested by \cite{Murray} and first used in practise by \cite{Strugnell, Hawley}.
The underlying idea is to maximise the likelihood of observing the combined kinematic observables of all sample stars (radial velocities and proper motions)
and their photometric distances by choosing the ``right'' combination of the kinematic parameters (the parameters of the bulk velocity field 
of the sample --- in the simplest case just the three components of the bulk velocity ---
and the components of the velocity dispersion tensor. A detailed description of the method can be found in the original book by \cite{Murray}
and in the papers by \cite{Hawley} and \cite{Rastorguev17}. However, maximum-likelihood estimators can be biased - e.g., in the simplest case
the variance estimator is known to be biased downward if the population mean is unknown \citep{Liu}. To reveal such biases, we generated for each solution
100 simulated data sets with stars fixed at the same sky positions as the stars of the actual sample and 
with the radial velocities and proper-motion components generated randomly (with the 1-d Gaussian distribution for radial velocities and 2-d Gaussian
distribution for proper motions) 
in accordance with the inferred velocity-field parameters
(in the simplest case just the components of the bulk mean velocity) and the components the velocity dispersion tensor plus the normally distributed 
radial-velocity and proper-motion errors. We also ``scattered'' the initial input star distances $D_i$ by adding normally distributed errors with zero mean
and with variance equal to 0.05~$D_i$ and multiplying the resulting distance values by the distance-scale correction factor $P$ inferred from the 
corresponding solution for the real data set. We then found the corresponding maximum-likelihood solutions for every such simulated set
and computed the mean values for all desired parameters, the differences between these mean values and the corresponding ``true'' (input) values --- those
given by the solution of the real set, and the standard deviations of  these differences. These differences provide an estimate of the eventual bias
of the real-data solution, and the standard deviations give us the estimates of the errors of the corresponding parameters, which we compare with the parameter 
errors given by the real-data solution.
}. Like in the case of our recent analysis of the kinematics of the RR Lyrae population \citep{Utkin18}, we assume that the halo is non-rotating
(while still determining the tangential velocity component that reflects the total angular momentum). { We address the possible rotation of our
halo sample in Section~\ref{rotation}}.  
Generally, for the entire
sample we aim to determine the following quantities:
(1) the velocity components of the sample relative to the Sun ($U_0$, $V_0$, $W_0$) in the Galactocentric Cartesian coordinate system:
$U_0$ in the direction toward the Galactic centre, $V_0$ in the direction of Galactic rotation, and $W_0$ in the direction toward the North Galactic Pole;
(2) the velocity dispersion components ($\sigma_r$, $\sigma_{\theta}$, $\sigma_{\phi}$) in the Galactocentric spherical coordinate system assuming
that the principal axes of the velocity ellipsoid are aligned with the local directions to the Galactic centre , Galactic rotation and that of the $\theta$
coordinate { (we address the possible deviation 
from this alignment in Section~\ref{misalignment})};
(3) the distance-scale correction factor $P$ such that the true distance $r_t$ is related to the adopted distance $r$ as $r_t$~=~$r/P$, and
(4) the solar Galactocentric distance $R_0$. 

{ We use the procedure adopted by \citet{Hawley} to estimate the standard errors in the inferred parameters. To this end,
we determine the error function $S$~=~-2~ln($LF$) and find the uncertainties in the final parameters by numerically computing its second derivatives at the
inferred minimum, $S_0$. To this end, we fix the particular parameter at its value at the minimum, $X_i(min)$, add a small term $d_i$ so that
$X_i$ = $X_i(min)$ + $d_i$ and
then allow other parameters to converge to a new minimum, $S1$. We finally estimate the variance in the inferred $X_i$  as

\begin{equation}
\sigma_i^2~=~d_i^2/(S_i-S_0)
\end{equation}
} We also compute the total velocity dispersion,
$\sigma V$~=~$\sqrt{\sigma_r^2 + \sigma_{\phi}^2 + \sigma_{\theta}^2}$ and the anisotropy parameter $\beta$ defined as:

\begin{equation}
\label{eq:beta_def}
\beta = 1 - \frac{\sigma_{\theta}^2 + \sigma_{\phi}^2}{2\sigma_r^2}.
\end{equation} 

To have a more detailed picture of the kinematics of our BHB star sample, we take advantage of the large number of objects involved 
and  analyse the variations of the velocity-field parameters with
Galactocentric distance. To this end, we subdivide the filtered sample into 2-kpc wide Galactocentric-distance bins spanning the interval from
3 to 23~kpc plus three broader bins for more distant stars (23 to 27~kpc, 27 to 35~kpc, and 35 to 60~kpc). To make the single-bin solutions more stable,
we fix the velocity components of the sample relative to the Sun ($U_0$, $V_0$, $W_0$) at their values determined for the entire sample and 
assume that the velocity ellipsoid is two-axial, i.e., the 
two axes perpendicular to the direction toward the Galactic centre are equal, and infer only  the 
velocity dispersion along Galacticentric radius, $\sigma_r$, and anisotropy parameter $\beta$ rather than estimating all three axes 
($\sigma_r$, $\sigma_{\theta}$, $\sigma_{\phi}$) of the ellipsoid.

\section{Results}
\label{sec:results}

\subsection{Bulk solution}

Table~\ref{all_xue} gives the results obtained by applying the maximum-likelihood statistical-parallax method to the entire 
decontaminated subsample of 2582 blue horizontal branch stars. { Column~1 gives the names of
 the inferred parameters; column~2, their
inferred optimum values; columns~3 and 4, the corresponding minimum and maximum values obtained by cross-sectioning the
likelihood-function profile near its global minimum by the hyperplane $LF$ = $LF_0$ + 1, where $LF_0$ 
is the minimum of the likelihood function as described above; column~5, the uncertainty, and column~6, the corresponding unit of measure.}

\begin{table}
  \centering
  \caption{The velocity-field parameters for the entire decontaminated BHB star sample with initial distances from \citet{XueBHB11}. }\label{all_xue}
\begin{tabular}{r r r r r r}
\hline
 Values            & opt & min & max & uncertainty & units\\
\hline
 $U_0$             &   -7.5 &   -9.3 &   -5.7 & 1.8      &  \\
 $V_0$             & -240.2 & -244.6 & -236.4 & 4.1      &  \\
 $W_0$             &   -5.4 &   -7.3 &   -3.5 & 1.9      &  \\
                   &        &        &        &          & km/s   \\
 $\sigma_r$        &  112.4 &  110.8 &  114.1 & 1.7      &  \\
 $\sigma_{\phi}$   &   77.6 &   76.0 &   79.6 & 1.8      &  \\
 $\sigma_{\theta}$ &   79.6 &   78.0 &   81.2 & 1.6      &  \\
 $\sigma V$        &  158.1 &  155.4 &  161.1 & 2.9      &  \\
 $\beta$           &  0.508 &  0.487 &  0.525 & 0.020    &  \\
\hline
 P                 &  1.051 & 1.032  & 1.068  & 0.017   &  \\
\hline
 $R_0$             &  8.23  & 7.82   & 9.00   & 0.59   & kpc   \\
\hline
\end{tabular}
\end{table}      

Table~\ref{all_xue_sim} summarises the results obtained by applying the maximum-likelihood statistical-parallax method to 100  data sets simulated
based on the parameter values obtained for decontaminated BHB star sample with initial star distances adopted from \cite{XueBHB11}. { Column~1 
gives the names of the inferred parameters; column~2, the mean values of these parameters averaged over 100 solutions for
simulated data sets; column~3, difference between this mean value and the input (``true'') value, which  ``measures'' the eventual bias
in the estimated of the corresponding parameter; column~4, the standard deviation of these differences, which serves as an estimate of
the standard error of the corresponding parameter; and column~5, the ratio of this standard deviation to the standard error of the 
corresponding parameter given by the real-data solution (Table~\ref{all_xue}), and column~6, the units of measure.} { As is evident from this
table, the bias is not significant in all cases and the solution error estimates recover very well the scatter of the parameter
values obtained by solving simulated sets and therefore we apply no bias corrections to our solutions}.

\begin{table}
  \centering
  \caption{The summary of the results obtained for 100 simulated data sets based on the values given by 
the real-data solution in Table~\ref{all_xue}. 
Here ``Mean'' is the mean value of parameter $Param$ averaged over the maximum-likelihood solutions for 100 simulated data sets;
``True'' is the input value $Param_0$ given by the real-data solution in Table~\ref{all_xue}; Difference is the difference $<Param>$~-~$Param_0$;
scatter is the standard deviation of $<Param>$~-~$Param_0$ ($\sigma (<Param>$~-~$Param_0$), and the $\frac{S}{U}$ is the ratio of the scatter
to the uncertainty of the corresponding parameter given by the real-data solution (Column~5 of Table~\ref{all_xue}). }\label{all_xue_sim}
\begin{tabular}{r r r r r r r}
\hline
 Values            & Mean    & ``True'' & Difference & Scatter & units & $\frac{S}{U}$\\
\hline
 $U_0$             &   -7.8 &   -7.5 &   -0.3~$\pm$~0.2 & 1.6      & & 0.9 \\
 $V_0$             & -239.8 & -240.2 &   +0.5~$\pm$~0.4 & 4.1      & & 1.0 \\
 $W_0$             &   -5.3 &   -5.4 &   +0.1~$\pm$~0.2 & 2.0      & & 1.1 \\
                   &        &        &                  &   & km/s &   \\
 $\sigma_r$        &  112.5 &  112.4 &   +0.1~$\pm$~0.2 & 1.8      & & 1.0 \\
 $\sigma_{\phi}$   &   77.4 &   77.6 &   -0.2~$\pm$~0.2 & 1.7      & & 1.0 \\
 $\sigma_{\theta}$ &   79.5 &   79.6 &   -0.1~$\pm$~0.2 & 1.8      & & 1.1 \\
 $\sigma V$        &  158.0 &  158.1 &   -0.1~$\pm$~0.2 & 2.3      & & 0.8 \\
 $\beta$           &  0.513 &  0.511 & +0.002~$\pm$~0.002 & 0.022  & & 1.1 \\
\hline
 P                 &  1.054 & 1.051  & +0.003~$\pm$~0.002  & 0.017   & & 1.0 \\
\hline
 $R_0$             &  8.15  & 8.23   & -0.08~$\pm$~0.06   & 0.63   & kpc & 1.1  \\
\hline
\end{tabular}
\end{table}      

Our estimate for the solar Galactocentric distance, $R_0$~=~8.2~$\pm$~0.6~kpc agrees well with most of the recent determinations 
of this parameter. { Thus the most precise and accurate $R_0$ determination based on the 16-year orbit of the star S2 
around the massive black hole Sgr A$^*$ measured astrometrically and spectroscopically for 27 years by the Gravity collaboration/
Galactic centre is $R_0$~=~8.178~$\pm$~0.013~$\pm$~0.022~kpc~\citep{Abuter}. In their comprehensive review, \cite{BlandHawthorn}
derive $R_0$~=~8.2~$\pm$~0.1~kpc~as their best estimate.} The authors of more recent reviews, e.g., \cite{Camarillo, Vallee} 
found the median of recent $R_0$ estimates to be of about $R_0$~=~8.0~$\pm$~0.3~kpc
and $R_0$~=~8.0~$\pm$~0.2~kpc, respectively, and an analysis of the kinematics of Galactic masers by \cite{Rastorguev17} yields
$R_0$~=~8.24~$\pm$~0.12~kpc. A recent analysis of the photometry of type-II Cepheids in the Galactic bulge yields
$R_0$~=~8.46~$\pm$~0.03~$\pm$~0.11~kpc \citep{Braga}, whereas the near-IR photometry of the RR Lyrae star population near the Galactic center 
yields $R_0$~=~8.05~$\pm$~0.02~kpc \citep{Contreras}, and an estimate based on globular-cluster kinematics  yields
$R_0$~=~7.6~$\pm$~0.7~kpc \citep{K18}. Our estimate of the mean velocity component in the direction of Galactic rotation,
$V_0$~=~-239~$\pm$~4~km/s, which can be viewed as the corresponding velocity component of the reflex solar motion with respect  to the 
Galactic rest frame, also agrees with other recent estimates of this quantity (e.g., $V_0$~=~-231.4~$\pm$~1.6~km/s from our analysis of the motions
of halo RR Lyrae type variables \citep{Utkin18}, $V_0$~=~-231~$\pm$~19~km/s from an analysis of the motions of metal-poor globular clusters \citep{K18},
$V_0$~=~-254~$\pm$~7~km/s from an analysis of Galactic maser motions).

Interestingly, our estimate of the correction factor to the BHB distance scale , $P$~=~1.051~$\pm$~0.017, implies that the 
BHB star distances given by \cite{XueBHB11} should be reduced  by a factor of 1.051 rather than increased by a factor of 1.06~$\pm$~0.03 found in our analysis published
a decade ago \citep{Dambis10} based on a smaller BHB star sample of \cite{XueBHB08} combined with SDSS proper motions \citep{SDSS7}. We, naturally,
believe the current estimate, which is based on a more extensive sample and much more accurate Gaia DR2 proper motions, to be  more reliable.

To test whether our results are sensitive to the metallicity dependence of the BHB star magnitudes, we repeated our computations with the BHB star
distances computed using the $M_g$ absolute-magnitude calibration proposed by \cite{Fermani2} (their equation~(5)). We provide the results obtained
with the  same cuts (all stars within the Sgr stream region, all stars within less than 5~kpc from the Galactic midplane, and all stars
with Galactic rest-frame
velocities higher than 600~km/s excluded) --- now 2607 objects ---
in Table~\ref{all_fermani}.

\begin{table}
  \centering
  \caption{The velocity-field parameters for the entire  decontaminated BHB star sample
leaving only objects farther than 5~kpc from the Galactic midplane with distances computed using
the metallicity-dependent calibration of \citet{Fermani2}. }\label{all_fermani}
\begin{tabular}{r r r r r r}
\hline
 Values            & opt & min & max & uncertainty & units\\
\hline
 $U_0$             &   -8.1 &   -9.9 &   -6.3 & 1.8      &  \\
 $V_0$             & -239.0 & -242.9 & -235.3 & 3.7      &  \\
 $W_0$             &   -6.1 &   -8.0 &   -4.2 & 1.9      &  \\
                   &        &        &        &          & km/s   \\
 $\sigma_r$        &  112.6 &  111.0 &  114.3 & 1.7      &  \\
 $\sigma_{\phi}$   &   77.8 &   76.2 &   79.7 & 1.8      &  \\
 $\sigma_{\theta}$ &   79.8 &   78.2 &   81.4 & 1.6      &  \\
 $\sigma V$        &  158.4 &  155.7 &  161.4 & 2.9      &  \\
 $\beta$           &  0.511 &  0.490 &  0.528 & 0.019    &  \\
\hline
 P                 &  1.084 & 1.066  & 1.102  & 0.018   &  \\
\hline
 $R_0$             &  8.09  & 7.70   & 8.66   & 0.48    & kpc   \\
\hline
\end{tabular}
\end{table}

As is evident from Table~\ref{all_fermani}, the main kinematical parameters and $R_0$ estimate 
remain practically the same as in the case
of the (metallicity independent) calibration used by \cite{XueBHB11}.

\begin{table}
  \centering
  \caption{The summary of the results obtained for 100 simulated data sets based on the values given by the real-data solution in Table~\ref{all_fermani}. 
Here ``Mean'' is the mean value of parameter $Param$ averaged over the maximum-likelihood solutions for 100 simulated data sets;
``True'' is the input value $Param_0$ given by the real-data solution in Table~\ref{all_fermani}; Difference is the difference $<Param>$~-~$Param_0$;
scatter is the standard deviation of $<Param>$~-~$Param_0$ ($\sigma (<Param>$~-~$Param_0$), and the $\frac{S}{U}$ is the ratio of the scatter
to the uncertainty of the corresponding parameter given by the real-data solution (Column~5 of Table~\ref{all_fermani}). }\label{all_fermani_sim}
\begin{tabular}{r r r r r r r}
\hline
 Values            & Mean    & ``True'' & Difference & Scatter & units & $\frac{S}{U}$\\
\hline
 $U_0$             &   -7.8 &   -8.1 &   +0.3~$\pm$~0.2 & 2.0      & & 1.1 \\
 $V_0$             & -238.7 & -239.0 &   +0.3~$\pm$~0.4 & 4.2      & & 1.1 \\
 $W_0$             &   -6.2 &   -6.1 &   -0.1~$\pm$~0.2 & 1.7      & & 0.9 \\
                   &        &        &                  &   & km/s &   \\
 $\sigma_r$        &  112.8 &  112.6 &   +0.2~$\pm$~0.2 & 1.6      & & 1.0 \\
 $\sigma_{\phi}$   &   77.6 &   77.8 &   -0.2~$\pm$~0.2 & 1.7      & & 1.0 \\
 $\sigma_{\theta}$ &   79.9 &   79.8 &   +0.1~$\pm$~0.2 & 1.9      & & 1.2 \\
 $\sigma V$        &  158.0 &  158.1 &   -0.1~$\pm$~0.2 & 2.3      & & 0.8 \\
 $\beta$           &  0.512 &  0.510 & +0.002~$\pm$~0.002 & 0.022  & & 1.1 \\
\hline
 P                 &  1.087 & 1.084  & +0.003~$\pm$~0.002  & 0.020   & & 1.1 \\
\hline
 $R_0$             &  8.10  & 8.09   & +0.01~$\pm$~0.07   & 0.68   & kpc & 1.4  \\
\hline
\end{tabular}
\end{table}

Table~\ref{all_fermani_sim} summarises the results obtained by applying the maximum-likelihood statistical-parallax method to 100  data sets simulated
based on the parameter values obtained for decontaminated BHB star sample with initial star distances computed using the metallicity-dependent
calibration of \cite{Fermani2}. The layout of this table is identical to that of Table~\ref{all_xue_sim}. As is evident from Tables~\ref{all_xue_sim} and \ref{all_fermani_sim},
 the bias is not significant in all cases and therefore we apply no bias corrections to our solutions. Furthermore,
the solution error estimates agree quite  well with the scatter of the parameter
values obtained by solving simulated sets.

\subsection{Deviation from Galactocentric alignment of the velocity ellipsoid}\label{misalignment}

Most of the studies find the velocity ellipsoid of halo stars to be close-to-spherically aligned \citep{Smith2009,Bond2010,Evans2016, Wegg2018,Everall} 
---
as we assume in our analysis. To explore the effect of the deviation from spherical alignment,
we use the following parametrisation of the spatial dependence of the tilt $\alpha$ (the tangent of the tilt angle) 
of the longest axis of the velocity ellipsoid with respect to the
Galactic midplane proposed by \citet{Binney2014} and \citet{Budenbender2015}: 
\begin{equation}
\alpha = \alpha_0\,\arctan{|z| / R},
\label{eq:alpha}
\end{equation}
where $\alpha$~=1.0 and $\alpha$~=0.0 correspond to strictly radial and cylindrical alignment, respectively. We computed a solution with $\alpha_0$~=~const
treated as an extra free parameter. The results are summarised in Tables~\ref{all_xue_tilt} and \ref{all_fermani_tilt}.

\begin{table}
  \centering
  \caption{The velocity-field parameters for the entire  decontaminated BHB star sample
leaving only objects farther than 5~kpc from the Galactic midplane with the initial distances adopted
from \citet{XueBHB11} with velocity-ellipsoid alignment parameter $\alpha_0$ treated as an extra free parameter. }\label{all_xue_tilt}
\begin{tabular}{r r r r r r}
\hline
 Values            & opt &  uncertainty & units\\
\hline
 $U_0$             &   -7.3  & 1.8      &  \\
 $V_0$             & -240.6  & 3.9      &  \\
 $W_0$             &   -5.5  & 1.9      &  \\
                   &         &          &  km/s   \\
 $\sigma_r$        &  112.5 & 1.6      &  \\
 $\sigma_{\phi}$   &   77.9 & 1.9      &  \\
 $\sigma_{\theta}$ &   78.9 & 1.6      &  \\
 $\sigma V$        &  157.9 & 1.7      &  \\
 $\beta$           &  0.515 & 0.023    &  \\
 $\alpha_0$        &  1.172 & 0.080    &  \\
\hline
 P                 &  1.052 & 0.018   &  \\
\hline
 $R_0$             &  8.70  & 0.74    & kpc   \\
\hline
\end{tabular}
\end{table}

\begin{table}
  \centering
  \caption{The velocity-field parameters for the entire  decontaminated BHB star sample
leaving only objects farther than 5~kpc from the Galactic midplane with the initial distances computed using
the metallicity-dependent calibration of \citet{Fermani2} with velocity-ellipsoid alignment parameter $\alpha_0$ treated as an extra free parameter. }\label{all_fermani_tilt}
\begin{tabular}{r r r r r r}
\hline
 Values            & opt &  uncertainty & units\\
\hline
 $U_0$             &   -7.9  & 1.8      &  \\
 $V_0$             & -239.1  & 4.1      &  \\
 $W_0$             &   -6.2  & 1.9      &  \\
                   &         &          & km/s   \\
 $\sigma_r$        &  112.7 & 1.7      &  \\
 $\sigma_{\phi}$   &   78.0 & 2.0      &  \\
 $\sigma_{\theta}$ &   79.0 & 1.6      &  \\
 $\sigma V$        &  158.2 & 1.6      &  \\
 $\beta$           &  0.515 & 0.022    &  \\
 $\alpha_0$        &  1.174 & 0.087    &  \\
\hline
 P                 &  1.087 & 0.025   &  \\
\hline
 $R_0$             &  8.49  & 0.76    & kpc   \\
\hline
\end{tabular}
\end{table}

We can see that allowing for deviation from spherical alignment of the velocity ellipsoid has practically no effect on all
the inferred parameters except $R_0$, which increases by $\sim$~0.4--0.5~kpc, i.e., by about one standard deviation.
Interestingly, our estimate of the parameter $\alpha_0$ is marginally greater than unity (by two standard deviations), in contrast
to the results of all other studies, which yield values between  0.0 and 1.0 (mostly close to $\alpha_0$~=1.0). Given that all other
parameters remain practically intact we set $\alpha_0$~=1.0 in all our subsequent computations (i.e., assume spherical alignment
of the velocity ellipsoid).

\subsection{Rotation of the sample }\label{rotation}
We now test our assumption that the decontaminated BHB star sample is nonrotating. To this end, we introduce the linear rotation velocity
$V_{rot}$, which we assume to be independent of the distance from the rotation axis (flat rotation curve):
$$
V_x = V_x(0) - (V_{rot}/R_G) y
$$

\begin{equation}
V_y = V_y(0) + (V_{rot}/R_G) x,
\end{equation}
where $R_G$ is the distance of the star from the Galactic rotation axis. The results are summarised in Tables~\ref{all_xue_rot} and \ref{all_fermani_rot}.

\begin{table}
  \centering
  \caption{The velocity-field parameters for the entire  decontaminated BHB star sample
leaving only objects farther than 5~kpc from the Galactic midplane with the initial distances adopted
from \citet{XueBHB11} with the fixed overall rotation velocity $V_{rot}$ (flat rotation) treated as an extra free parameter. }\label{all_xue_rot}
\begin{tabular}{r r r r r r}
\hline
 Values            & opt &  uncertainty & units\\
\hline
$U_0$             &   -7.1  & 1.9      &  \\
 $V_0$             & -240.6  & 4.2      &  \\
 $W_0$             &   -5.5  & 1.9      &  \\
                   &         &          &  km/s   \\
 $\sigma_r$        &  112.3 & 1.7      &  \\
 $\sigma_{\phi}$   &   77.4 & 1.7      &  \\
 $\sigma_{\theta}$ &   79.5 & 1.6      &  \\
 $\sigma V$        &  157.9 & 1.7      &  \\
 $\beta$           &  0.512 & 0.024    &  \\
 $V_{rot}$         &  1.6   & 2.1      &  \\
\hline
 P                 &  1.053 & 0.020   &  \\
\hline
 $R_0$             &  8.18  & 0.41    & kpc   \\
\hline
\end{tabular}
\end{table}

\begin{table}
  \centering
  \caption{The velocity-field parameters for the entire  decontaminated BHB star sample
leaving only objects farther than 5~kpc from the Galactic midplane with the initial distances computed using
the metallicity-dependent calibration of \citet{Fermani2} with fixed overall rotation velocity $V_{rot}$ (flat rotation) 
 treated as an extra free parameter. }\label{all_fermani_rot}
\begin{tabular}{r r r r r r}
\hline
 Values            & opt &  uncertainty & units\\
\hline
 $U_0$             &   -7.6  & 1.9      &  \\
 $V_0$             & -239.5  & 3.9      &  \\
 $W_0$             &   -6.1  & 1.9      &  \\
                   &         &          &  km/s   \\
 $\sigma_r$        &  112.6 & 1.6      &  \\
 $\sigma_{\phi}$   &   77.7 & 1.7      &  \\
 $\sigma_{\theta}$ &   79.7 & 1.6      &  \\
 $\sigma V$        &  158.3 & 1.7      &  \\
 $\beta$           &  0.511 & 0.024    &  \\
 $V_{rot}$         &  1.8   & 2.1      &  \\
\hline
 P                 &  1.086 & 0.019   &  \\
\hline
 $R_0$             &  8.05  & 0.40    & kpc   \\
\hline
\end{tabular}
\end{table}

We conclude that overall rotation of the sample is negligible 
and statistically insignificant ($V_{rot}$~$\sim$~2~$\pm$~2~km/s)  and we therefore ignore it in subsequent computations. { Our estimate for the
overall halo rotation velocity agrees well with the estimates obtained by \citet{Bajkova} (1~$\pm$~4~km/s) and \citet{K18} (-17~$\pm$~17~km/s) 
based on the kinematics of globular clusters, and is slightly inconsistent with the estimate by \citet{Deason17} (14~$\pm$~2~$\pm$~10~km/s)
based on the kinematic data for RR Lyrae, blue horizontal branch stars, and K giant stars with pre-Gaia-DR2 proper motions, and is at variance
with the estimate of \citet{Tian} ((+27$^{+4}_{-5}$~km/s)) based on the kinematics of a sample of metal-poor K-type giants. However, the latter
tracers can be contaminated by thick-disk stars of the same type. Our result also agrees with the halo rotation estimate by 
\citet{Kafle17} (-7~$\pm$~8~km/s) 
based on metal-poor K-type giants, but is at variance with another estimate obtained in the same study (26~$\pm$~4~km/s) based on main-sequence turnoff stars.}

\subsection{Systematic error in SDSS radial velocities}
Possible systematics in SDSS radial-velocity errors may also affect the inferred kinematic parameters as well as distance-scale correction factor and
$R_0$. To test the extent of this effect, we incorporate a systematic shift in radial velocities $\Delta V_r$ (in the sense $V_r$(true)~=~$V_r$(SDSS)~+~$\Delta V_r$)
into our model and compute the corresponding solution. The results are summarised in Tables~\ref{all_xue_rv_error} and \ref{all_fermani_rv_error}.

\begin{table}
  \centering
  \caption{The velocity-field parameters for the entire  decontaminated BHB star sample
leaving only objects farther than 5~kpc from the Galactic midplane with the initial distances adopted
from \citet{XueBHB11} with systematic radial-velocity offset $\Delta V_r$ (in the sense $V_r$(true)~=~$V_r$(SDSS)~+~$\Delta V_r$) 
treated as an extra free parameter. }\label{all_xue_rv_error}
\begin{tabular}{r r r r r r}
\hline
 Values            & opt &  uncertainty & units\\
\hline
 $U_0$             &   -6.5  & 1.9      &  \\
 $V_0$             & -237.3  & 4.2      &  \\
 $W_0$             &   -2.9  & 1.9      &  \\
                   &         &          &  km/s   \\
 $\sigma_r$        &  112.0 & 1.7      &  \\
 $\sigma_{\phi}$   &   76.8 & 1.7      &  \\
 $\sigma_{\theta}$ &   78.9 & 1.6      &  \\
 $\sigma V$        &  157.1 & 1.7      &  \\
 $\beta$           &  0.517 & 0.024    &  \\
 $\Delta V_r$      &  +5.2  & 2.6      &  \\
\hline
 P                 &  1.063 & 0.017   &  \\
\hline
 $R_0$             &  8.04  & 0.56    & kpc   \\
\hline
\end{tabular}
\end{table}

\begin{table}
  \centering
  \caption{The velocity-field parameters for the entire  decontaminated BHB star sample
leaving only objects farther than 5~kpc from the Galactic midplane with the initial distances computed using
the metallicity-dependent calibration of \citet{Fermani2} with systematic radial-velocity offset 
$\Delta V_r$ (in the sense $V_r$(true)~=~$V_r$(SDSS)~+~$\Delta V_r$) 
treated as an extra free parameter. }\label{all_fermani_rv_error}
\begin{tabular}{r r r r r r}
\hline
 Values            & opt &  uncertainty & units\\
\hline
 $U_0$             &   -7.1  & 1.8      &  \\
 $V_0$             & -236.0  & 4.1      &  \\
 $W_0$             &   -3.6  & 2.2      &  \\
                   &         &          &  km/s   \\
 $\sigma_r$        &  112.2 & 1.7      &  \\
 $\sigma_{\phi}$   &   77.0 & 1.0      &  \\
 $\sigma_{\theta}$ &   79.1 & 1.6      &  \\
 $\sigma V$        &  157.4 & 1.7      &  \\
 $\beta$           &  0.516 & 0.025    &  \\
 $\Delta V_r$      &  +5.3  & 2.6      &  \\
\hline
 P                 &  1.097 & 0.020   &  \\
\hline
 $R_0$             &  7.91  & 0.48    & kpc   \\
\hline
\end{tabular}
\end{table}      

As is evident from a comparison of Tables~\ref{all_xue_rv_error} and \ref{all_fermani_rv_error} with Tables~\ref{all_xue} and \ref{all_fermani}, 
the radial-velocity offset 
is slightly significant (at the 2$\sigma$ level) in the sense that SDSSradial velocities are, on the average, smaller by $\sim$~5~km/s, and
this offset has only a marginal effect on other inferred parameters decreasing slightly the $R_0$ estimate, which still remains quite consistent
with most of the recent determinations. 

{
\subsection{Systematic error in proper motions}
To assess the errors due to systematic errors in Gaia DR2 proper motions, we added to Gaia DR2 proper motions the corrections proposed by 
\citet{Lindegren} for bright stars ($G<$~12.0):
\begin{equation}
\Delta pm_{\alpha}~=~\omega_x sin(\delta)cos(\alpha) + \omega_y sin(\delta) cos(\alpha) - \omega_z cos(\delta)
\label{pma_systematic}
\end{equation}
and
\begin{equation}
\Delta pm_{\delta}~=~\omega_x sin(\alpha)+ \omega_y cos(\alpha)
\label{pmd_systematic}
\end{equation}
where $\omega_x$~=~0.086~$\pm$~0.025~mas/yr, $\omega_y$~=~0.114~$\pm$~0.025~mas/yr, and $\omega_z$~=~0.037~$\pm$~0.025~mas/yr.
These corrections by all means far exceed the actual systematic errors of Gaia DR2 proper motions for stars of our sample, which are much fainter 
(see the histogram in Fig.~\ref{g_mag}) and most likely hardly need any systematic corrections \citep{Lindegren}. The results obtained with these corrections 
applied differ from those computed with raw Gaia DR2 proper motions (Figs.~(\ref{all_fermani}) and (\ref{all_fermani})) by  
$\Delta U_0$~=~+2.3~km/s, $\Delta V_0$~=~-0.05~km/s, $\Delta W_0$~=~-1.5~km/s, $\Delta \sigma_r$~=~+0.4~km/s, $\Delta \sigma_{\phi}$~=~+0.8~km/s,
$\Delta \sigma_{\theta}$~=~+0.8~km/s, $\Delta \sigma_V$~=~+1.1~km/s, $\Delta \beta$~=~-0.005, $\Delta P$~=~+0.016, and $\Delta R_0$~=~+0.23~kpc.
Thus the average maximum extra rms errors are of about 1.6~km/s for bulk velocity  parameters and 1.0~km/s for velocity-dispersion parameters increasing the
uncertainties of the former by a factor of 1.1--1.3 and those of the  latter by a factor of 1.1--1.2. The errors of the inferred 
parameters $\beta$, $P$, and $R_0$ increase by a maximum of a factor of 1.4, 1.3, and 1.1, respectively. However, the actual error increases
must be much smaller (according to \citet{Lindegren}, ``For G = 13 to 16 there are very few
comparison data but probably no correction is needed in that interval'' and the same appears to be true at fainter magnitudes) 
and perhaps hardly noticeable.
 }

{ 
\subsection{Halo substructures}
According to recent results,  the kinematic behaviour of the stellar halo is to a significant extent influenced by some major accretion events 
\citep{Ibata94, Helmi, Belokurov, Myeong, Koppelman, Belokurov20, Naidu}. 
However,  this substructure should not have had an important effect on our results. Given the 
data provided by
\citet{Naidu}, the structures Aleph, in-situ halo, and high-$\alpha$ disk can make up for a maximum of 3.5\% of our sample because 
of the respective metallicity distributions. In any case, imposing a [Fe/H]$<$~-1.0 cut totally eliminates these features
changing the inferred parameter values by less than 0.4~$\sigma$ (-0.2, +1.5, -0.5, +0.4, -0.8, and +0.7~km/s 
for $U_0$, $V_0$, $W_0$, $\sigma_r$, $\sigma_{\phi}$, and $\sigma_{\theta}$, respectively, -0.004 for scale factor $P$,
and -0.07~kpc for $R_0$).  The total fraction of Thamnos, Wukong, Milky-Way Thick Disk, 
Arjuna, Sequoia, and l'Itoi, and Helmi stream features account for 9\% of the halo population above 2 kpc, but are appreciably less
 represented at heights above 6 kpc. 
The major possible contaminant - the Sagittarius stream --
is excluded via position masking. This leaves only 
the Gaia-Enceladus Sausage as the dominant feature in our sample plus a small fraction of unclassified halo-like debris, both
with small $L_z$ component of angular momentum, resulting
in a radially dependent non-Gaussian velocity distribution. However, \cite{Popowski} showed the statistical parallax method 
to be extremely robust 
against particular form the velocity distribution and its deviations  from Gaussian and therefore we expect the main results reported 
here  not not be substantially  influenced by this factor.
}

\subsection{Dependence of kinematic parameters on Galactocentric distance}

We summarise  the results obtained by applying the statistical-parallax method to the
subsamples of the decontaminated sample limited by Galactocentric distance $R_G$ in Tables~\ref{xue_bin_uvwfixed}, 
\ref{xue_bin_uwfixed}, and~\ref{xue_bin_uvwfree} .
Table~\ref{xue_bin_uvwfixed} lists the kinematic parameters obtained with all average velocity components --- $U_0$, $V_0$, and $W_0$ --
fixed at their values inferred for the entire decontaminated sample (Table~\ref{all_xue}), whereas Table~\ref{xue_bin_uwfixed} gives the
solutions obtained with $V_0$ treated as a free parameter, and Table~\ref{xue_bin_uvwfree} gives the solutions
obtained with $U_0$, $V_0$, and $W_0$ treated as free parameters. In Table~\ref{xue_bin_uvwfixed}  column~1 ($R_G$ bin) 
gives the interval
of Galactocentric distances;  column~2, the number of stars in the bin; column~3, the average Galactocentric distance, $<R_G>$;
column~4, the radial velocity dispersion component, $\sigma_r$ with its standard error, column~5, the total 
velocity dispersion component, $\sigma V$ with its standard error. The last two columns (Columns~6 and 7 ) 
 give the anisotropy parameter $\beta$ and the distance-scale correction factor $P$, respectively,
with their standard errors.  In Table~\ref{xue_bin_uwfixed}  column~1 ($R_G$ bin) 
gives the interval
of Galactocentric distances;  column~2, the number of stars in the bin; column~3, the average Galactocentric distance, $<R_G>$;
column~4, the radial velocity dispersion component, $\sigma_r$ with its standard error, and column~5, the total 
velocity dispersion component, $\sigma V$ with its standard error. Column~7 gives the average velocity component $V_0$
in the direction of Galactic rotation, and columns~8 amd 9 give the anisotropy parameter $\beta$ and the 
distance-scale correction factor $P$, respectively,
with their standard errors. In Table~\ref{xue_bin_uvwfree}  column~1 ($R_G$ bin) 
gives the interval
of Galactocentric distances;  column~2, the number of stars in the bin; column~3, the average Galactocentric distance, $<R_G>$;
column~4, the radial velocity dispersion component, $\sigma_r$ with its standard error, and column~5, the total 
velocity dispersion component, $\sigma V$ with its standard error. Columns~7, 8, and 9 give the average velocity components $U_0$,
$V_0$, and $W_0$ in the direction toward the Galactic center, in the direction of Galactic rotation, and in the
direction toward the North Galactic Pole, respectively. Columns~10 and 11 give the anisotropy parameter $\beta$ and the 
distance-scale correction factor $P$, respectively,
with their standard errors.

\begin{table*}
  \centering
  \caption{The velocity-field parameters for the Galactocentric-distance binned subsamples
  (initial distances adopted from  \citet{XueBHB11})
with $U_0$, $V_0$, and $W_0$ fixed at the values listed in Table~\ref{all_xue}.}\label{xue_bin_uvwfixed}
{%
\begin{tabular}{r r r r r r r}
\hline
 $R_G$ bin   &  N   & $<R_G>$  & $\sigma_r$ & $\sigma V$ & $\beta$ & $P$  \\
             &      &          & km/s       & km/s       &         &      \\
\hline
   5--9           &  181    &  7.7               & 121.2~$\pm$~6.9   & 171.3~$\pm$~11.3 & 0.501~$\pm$~0.062 & 1.134~$\pm$~0.028  \\
  9--11           &  194    & 10.1               & 130.4~$\pm$~6.9   & 175.6~$\pm$~ 9.9 & 0.593~$\pm$~0.048 & 1.040~$\pm$~0.024  \\
 11--13           &  258    & 12.0               & 128.5~$\pm$~5.8   & 165.0~$\pm$~ 8.4 & 0.676~$\pm$~0.033 & 1.076~$\pm$~0.020  \\
 13--15           &  244    & 14.1               & 126.8~$\pm$~5.9   & 176.8~$\pm$~ 9.6 & 0.528~$\pm$~0.051 & 1.047~$\pm$~0.024  \\
 15--17           &  246    & 16.0               & 111.3~$\pm$~5.1   & 154.6~$\pm$~ 7.8 & 0.535~$\pm$~0.051 & 1.046~$\pm$~0.022  \\
 17--19           &  241    & 17.9               & 112.8~$\pm$~5.3   & 156.7~$\pm$~ 8.3 & 0.535~$\pm$~0.052 & 1.023~$\pm$~0.022  \\
 19--21           &  209    & 20.0               & 105.6~$\pm$~5.3   & 154.5~$\pm$~ 8.8 & 0.430~$\pm$~0.070 & 1.028~$\pm$~0.026  \\
 21--23           &  187    & 22.0               &  99.6~$\pm$~5.3   & 143.2~$\pm$~ 9.6 & 0.467~$\pm$~0.069 & 1.032~$\pm$~0.025  \\
 23--27           &  282    & 24.8               & 102.6~$\pm$~4.4   & 143.3~$\pm$~ 7.1 & 0.524~$\pm$~0.051 & 1.050~$\pm$~0.022  \\
 27--35           &  239    & 30.7               &  96.4~$\pm$~4.6   & 150.1~$\pm$~ 8.9 & 0.287~$\pm$~0.090 & 1.019~$\pm$~0.027  \\
 35--60           &  301    & 43.2               &  93.7~$\pm$~4.0   & 148.5~$\pm$~ 9.8 & 0.244~$\pm$~0.108 & 1.014~$\pm$~0.029  \\
\hline
\end{tabular}}
\end{table*}    

\begin{table*}
  \centering
  \caption{The velocity-field parameters for the Galactocentric-distance binned subsamples  (initial distances adopted from \citet{XueBHB11})
with $U_0$ and  $W_0$ fixed
at the values listed in Table~\ref{all_xue} and $V_0$ treated as a free parameter. }\label{xue_bin_uwfixed}
{%
\begin{tabular}{r r r r r r r r}
\hline
 $R_G$ bin   &  N   & $<R_G>$  & $\sigma_r$ & $\sigma V$ & $V_0$ & $\beta$ & $P$  \\
             &      &          & km/s       & km/s       & km/s  &         &      \\
\hline
   5--9           &  181    &  7.7                & 121.4~$\pm$~7.1   & 171.6~$\pm$~11.6 & -241.4~$\pm$~10.1  & 0.501~$\pm$~0.062 & 1.131~$\pm$~0.041  \\
  9--11           &  194    & 10.1                & 132.3~$\pm$~7.2   & 178.7~$\pm$~11.2 & -252.5~$\pm$~ 9.8  & 0.588~$\pm$~0.049 & 1.009~$\pm$~0.034  \\
 11--13           &  258    & 12.0                & 128.7~$\pm$~5.9   & 165.4~$\pm$~ 8.1 & -242.4~$\pm$~ 7.7  & 0.674~$\pm$~0.034 & 1.070~$\pm$~0.030  \\
 13--15           &  244    & 14.1                & 128.3~$\pm$~6.1   & 180.9~$\pm$~ 9.7 & -254.5~$\pm$~10.4  & 0.506~$\pm$~0.056 & 1.004~$\pm$~0.039  \\
 15--17           &  246    & 16.0                & 110.7~$\pm$~5.2   & 152.7~$\pm$~ 8.5 & -233.6~$\pm$~ 9.2  & 0.549~$\pm$~0.052 & 1.070~$\pm$~0.041  \\
 17--19           &  241    & 17.9                & 114.4~$\pm$~5.4   & 163.6~$\pm$~ 9.5 & -265.7~$\pm$~11.5  & 0.478~$\pm$~0.064 & 0.940~$\pm$~0.041  \\
 19--21           &  209    & 20.0                & 105.5~$\pm$~5.3   & 150.5~$\pm$~ 9.8 & -227.5~$\pm$~12.2  & 0.483~$\pm$~0.078 & 1.080~$\pm$~0.059  \\
 21--23           &  187    & 22.0                &  99.7~$\pm$~5.3   & 144.8~$\pm$~10.4 & -245.9~$\pm$~13.6  & 0.446~$\pm$~0.088 & 1.010~$\pm$~0.058  \\
 23--27           &  282    & 24.8                & 102.6~$\pm$~4.4   & 143.1~$\pm$~ 7.8 & -239.6~$\pm$~11.3  & 0.527~$\pm$~0.065 & 1.053~$\pm$~0.052  \\
 27--35           &  239    & 30.7                &  96.3~$\pm$~4.6   & 149.3~$\pm$~10.2 & -237.9~$\pm$~13.1  & 0.298~$\pm$~0.107 & 1.029~$\pm$~0.061  \\
 35--60           &  301    & 43.5                &  93.5~$\pm$~4.0   & 144.8~$\pm$~ 9.4 & -229.7~$\pm$~13.3  & 0.301~$\pm$~0.122 & 1.058~$\pm$~0.069  \\
\hline
\end{tabular}}
\end{table*} 

\begin{table*}
  \centering
  \caption{The velocity-field parameters for the Galactocentric-distance binned subsamples (initial distances adopted from \citet{XueBHB11})
with $U_0$, $V_0$, and  $W_0$  
treated as a free parameters. }\label{xue_bin_uvwfree}
{%
\begin{tabular}{r r r r r r r r r r}
\hline
 $R_G$ bin   &  N   & $<R_G>$  & $\sigma_r$ & $\sigma V$ & $U_0$ & $V_0$ & $W_0$ & $\beta$ & $P$  \\
             &      &          & km/s       & km/s       &         &      \\
\hline
   5--9           &  181    &  7.7                & 121.5~$\pm$~7.3   &172.1~$\pm$~12.1 &  -0.4~$\pm$~7.2  & -244.2~$\pm$~10.6   & -15.0~$\pm$~8.0  & 0.497~$\pm$~0.062 & 1.122~$\pm$~0.043  \\
  9--11           &  194    & 10.1                & 131.5~$\pm$~7.1   &177.2~$\pm$~10.5 & -14.0~$\pm$~6.7  & -250.1~$\pm$~ 9.9   & -11.3~$\pm$~7.1  & 0.592~$\pm$~0.049 & 1.021~$\pm$~0.035  \\
 11--13           &  258    & 12.0                & 129.0~$\pm$~6.0   &165.2~$\pm$~ 7.9 &  -7.9~$\pm$~5.3  & -241.4~$\pm$~ 7.8   &  +7.4~$\pm$~5.8  & 0.680~$\pm$~0.033 & 1.071~$\pm$~0.030  \\
 13--15           &  244    & 14.1                & 128.8~$\pm$~6.1   &181.4~$\pm$~ 9.5 &  +1.5~$\pm$~6.4  & -255.7~$\pm$~10.4   &  +1.5~$\pm$~6.4  & 0.508~$\pm$~0.056 & 0.998~$\pm$~0.039  \\
 15--17           &  246    & 16.0                & 110.5~$\pm$~5.2   &152.6~$\pm$~ 8.7 &  -6.4~$\pm$~5.4  & -233.8~$\pm$~ 9.3   &  +0.5~$\pm$~5.7  & 0.546~$\pm$~0.053 & 1.069~$\pm$~0.042  \\
 17--19           &  241    & 18.0                & 114.6~$\pm$~5.4   &163.1~$\pm$~ 9.4 &  +6.2~$\pm$~5.9  & -266.3~$\pm$~11.4   &  +1.0~$\pm$~6.2  & 0.487~$\pm$~0.063 & 0.937~$\pm$~0.041  \\
 19--21           &  209    & 20.1                & 105.5~$\pm$~5.3   &150.0~$\pm$~ 9.5 &  -7.5~$\pm$~5.9  & -226.3~$\pm$~12.3   &  -1.2~$\pm$~6.2  & 0.489~$\pm$~0.078 & 1.086~$\pm$~0.060  \\
 21--23           &  187    & 22.1                &  99.7~$\pm$~5.3   &144.6~$\pm$~ 9.3 &  -9.7~$\pm$~6.1  & -246.1~$\pm$~13.6   & -10.1~$\pm$~6.4  & 0.448~$\pm$~0.087 & 1.010~$\pm$~0.059  \\
 23--27           &  282    & 24.9                & 102.4~$\pm$~4.4   &143.4~$\pm$~ 7.7 &  +3.4~$\pm$~5.0  & -243.1~$\pm$~11.1   & -18.4~$\pm$~5.2  & 0.519~$\pm$~0.065 & 1.036~$\pm$~0.051  \\
 27--35           &  239    & 30.5                &  96.1~$\pm$~4.6   &149.8~$\pm$~10.4 & -14.4~$\pm$~6.1  & -240.1~$\pm$~13.6   & -12.7~$\pm$~6.3  & 0.285~$\pm$~0.113 & 1.018~$\pm$~0.063  \\
 35--60           &  301    & 44.5                &  93.0~$\pm$~6.1   &141.0~$\pm$~11.9 & -38.7~$\pm$~6.3  & -228.9~$\pm$~14.0   &  -9.7~$\pm$~6.1  & 0.351~$\pm$~0.122 & 1.070~$\pm$~0.074  \\
\hline
\end{tabular}}
\end{table*}

The results obtained by applying the statistical-parallax method to the
subsamples of the decontaminated sample limited by Galactocentric distance $R_G$ with initial distances computed using
the metallicity-dependent calibration of \citet{Fermani2} are presented in Tables~\ref{fermani_bin_uvwfixed}, 
\ref{fermani_bin_uwfixed}, and~\ref{fermani_bin_uvwfree} .
Table~\ref{fermani_bin_uvwfixed} lists the kinematic parameters obtained with all average velocity components --- $U_0$, $V_0$, and $W_0$ --
fixed at their values inferred for the entire decontaminated sample (Table~\ref{all_fermani}), whereas 
Table~\ref{fermani_bin_uwfixed} gives the
solutions obtained with $V_0$ treated as a free parameter, and Table~\ref{fermani_bin_uvwfree} gives the solutions
obtained with $U_0$, $V_0$, and $W_0$ treated as free parameters. Their layout is identical to that of Tables~\ref{xue_bin_uvwfixed}, 
\ref{xue_bin_uwfixed}, and~\ref{xue_bin_uvwfree}, respectively.

\begin{table*}
  \centering
  \caption{The velocity-field parameters for the Galactocentric-distance binned subsamples (initial distances computed using
the metallicity-dependent calibration of \citet{Fermani2})
with $U_0$, $V_0$, and $W_0$ fixed
at the values listed in Table~\ref{all_xue}. }\label{fermani_bin_uvwfixed}
{%
\begin{tabular}{r r r r r r r}
\hline
 $R_G$ bin   &  N   & $<R_G>$  & $\sigma_r$ & $\sigma V$ & $\beta$ & $P$  \\
             &      &          & km/s       & km/s       &         &      \\
\hline
   5--9           &  181    &  7.7               & 119.1~$\pm$~6.7   &  165.7~$\pm$~10.4  & 0.532~$\pm$~0.057 & 1.156~$\pm$~0.027  \\
  9--11           &  194    & 10.1               & 133.6~$\pm$~7.4   &  180.1~$\pm$~12.1  & 0.591~$\pm$~0.051 & 1.064~$\pm$~0.026  \\
 11--13           &  258    & 12.0               & 125.6~$\pm$~5.8   &  164.8~$\pm$~ 8.0  & 0.639~$\pm$~0.037 & 1.123~$\pm$~0.022  \\
 13--15           &  244    & 14.1               & 125.5~$\pm$~6.0   &  170.2~$\pm$~ 9.3  & 0.580~$\pm$~0.046 & 1.072~$\pm$~0.024  \\
 15--17           &  246    & 16.0               & 117.7~$\pm$~5.3   &  163.4~$\pm$~ 8.7  & 0.536~$\pm$~0.049 & 1.077~$\pm$~0.023  \\
 17--19           &  241    & 17.9               & 114.5~$\pm$~5.4   &  160.4~$\pm$~ 8.6  & 0.519~$\pm$~0.055 & 1.060~$\pm$~0.025  \\
 19--21           &  209    & 20.0               & 105.7~$\pm$~5.3   &  150.4~$\pm$~ 8.9  & 0.488~$\pm$~0.063 & 1.046~$\pm$~0.025  \\
 21--23           &  187    & 22.0               & 102.1~$\pm$~5.2   &  151.7~$\pm$~ 9.7  & 0.396~$\pm$~0.077 & 1.083~$\pm$~0.028  \\
 23--27           &  282    & 24.8               & 100.9~$\pm$~4.5   &  142.1~$\pm$~ 7.6  & 0.508~$\pm$~0.054 & 1.074~$\pm$~0.022  \\
 27--35           &  239    & 30.7               & 102.2~$\pm$~4.6   &  152.1~$\pm$~ 9.3  & 0.392~$\pm$~0.072 & 1.062~$\pm$~0.026  \\
 35--60           &  301    & 43.2               &  92.3~$\pm$~3.8   &  150.9~$\pm$~ 8.9  & 0.163~$\pm$~0.112 & 1.064~$\pm$~0.030  \\
\hline
\end{tabular}}
\end{table*}

\begin{table*}
  \centering
  \caption{The velocity-field parameters for the Galactocentric-distance binned subsamples  (initial distances computed using
the metallicity-dependent calibration of \citet{Fermani2})
with $U_0$ and  $W_0$ fixed
at the values listed in Table~\ref{all_xue} and $V_0$ treated as a free parameter. }\label{fermani_bin_uwfixed}
{%
\begin{tabular}{r r r r r r r r}
\hline
 $R_G$ bin   &  N   & $<R_G>$  & $\sigma_r$ & $\sigma V$ & $V_0$ & $\beta$ & $P$  \\
             &      &          & km/s       & km/s       & km/s  &         &      \\
\hline
   5--9           &  185    &  7.7                & 118.8~$\pm$~7.0   &  164.7~$\pm$~11.3  & -237.7~$\pm$~ 9.6  & 0.539~$\pm$~0.057 & 1.160~$\pm$~0.041  \\
  9--11           &  179    & 10.1                & 134.8~$\pm$~7.5   &  182.6~$\pm$~11.6  & -249.0~$\pm$~10.2  & 0.583~$\pm$~0.052 & 1.038~$\pm$~0.037  \\
 11--13           &  255    & 12.0                & 125.8~$\pm$~5.9   &  165.3~$\pm$~ 8.6  & -240.8~$\pm$~ 7.9  & 0.637~$\pm$~0.038 & 1.118~$\pm$~0.033  \\
 13--15           &  231    & 14.1                & 127.6~$\pm$~6.2   &  174.7~$\pm$~ 9.8  & -251.1~$\pm$~ 9.9  & 0.563~$\pm$~0.050 & 1.035~$\pm$~0.038  \\
 15--17           &  261    & 16.0                & 117.9~$\pm$~5.4   &  164.0~$\pm$~ 8.9  & -240.6~$\pm$~ 9.6  & 0.533~$\pm$~0.052 & 1.071~$\pm$~0.042  \\
 17--19           &  231    & 17.9                & 115.4~$\pm$~5.5   &  164.7~$\pm$~ 9.4  & -254.4~$\pm$~11.2  & 0.481~$\pm$~0.065 & 1.005~$\pm$~0.045  \\
 19--21           &  207    & 20.0                & 105.6~$\pm$~5.3   &  152.0~$\pm$~ 9.9  & -244.7~$\pm$~13.3  & 0.464~$\pm$~0.078 & 1.024~$\pm$~0.057  \\
 21--23           &  201    & 22.0                & 102.0~$\pm$~5.2   &  148.1~$\pm$~ 9.2  & -227.8~$\pm$~13.1  & 0.446~$\pm$~0.078 & 1.133~$\pm$~0.067  \\
 23--27           &  271    & 24.8                & 101.0~$\pm$~4.4   &  145.2~$\pm$~ 7.4  & -249.9~$\pm$~11.6  & 0.467~$\pm$~0.065 & 1.030~$\pm$~0.051  \\
 27--35           &  263    & 30.7                & 102.0~$\pm$~4.6   &  149.8~$\pm$~ 8.5  & -232.0~$\pm$~12.5  & 0.422~$\pm$~0.085 & 1.092~$\pm$~0.063  \\
 35--60           &  301    & 43.5                &  92.2~$\pm$~3.8   &  148.2~$\pm$~ 9.6  & -231.4~$\pm$~12.6  & 0.209~$\pm$~0.130 & 1.097~$\pm$~0.067  \\
\hline
\end{tabular}}
\end{table*}

\begin{table*}
  \centering
  \caption{The velocity-field parameters for the Galactocentric-distance binned subsamples (initial distances computed using
the metallicity-dependent calibration of \citet{Fermani2})
with $U_0$, $V_0$, and  $W_0$  
treated as a free parameters. }\label{fermani_bin_uvwfree}
{%
\begin{tabular}{r r r r r r r r r r}
\hline
 $R_G$ bin   &  N   & $<R_G>$  & $\sigma_r$ & $\sigma V$ & $U_0$ & $V_0$ & $W_0$ & $\beta$ & $P$  \\
             &      &          & km/s       & km/s       &         &      \\
\hline
   5--9           &  185    &  7.7                & 119.9~$\pm$~7.2   &  166.4~$\pm$~11.0  &  -0.9~$\pm$~6.9  & -241.6~$\pm$~10.1   & -14.6~$\pm$~7.6  & 0.537~$\pm$~0.057 & 1.146~$\pm$~0.043  \\
  9--11           &  179    & 10.1                & 133.9~$\pm$~7.5   &  180.9~$\pm$~11.7  & -15.1~$\pm$~7.2  & -247.4~$\pm$~10.3   & -18.1~$\pm$~8.0  & 0.587~$\pm$~0.052 & 1.049~$\pm$~0.038  \\
 11--13           &  255    & 12.0                & 126.0~$\pm$~5.9   &  165.3~$\pm$~ 8.3  &  -8.5~$\pm$~5.5  & -240.9~$\pm$~ 8.1   &  +5.2~$\pm$~5.9  & 0.639~$\pm$~0.038 & 1.114~$\pm$~0.033  \\
 13--15           &  231    & 14.1                & 127.2~$\pm$~6.2   &  173.9~$\pm$~ 9.9  &  -5.7~$\pm$~6.2  & -251.7~$\pm$~10.0   &  +2.7~$\pm$~6.6  & 0.566~$\pm$~0.050 & 1.029~$\pm$~0.038  \\
 15--17           &  261    & 16.0                & 117.8~$\pm$~5.4   &  164.2~$\pm$~ 8.4  &   1.6~$\pm$~5.7  & -242.5~$\pm$~ 9.6   &  -8.1~$\pm$~5.9  & 0.528~$\pm$~0.053 & 1.062~$\pm$~0.041  \\
 17--19           &  231    & 18.0                & 114.9~$\pm$~5.5   &  163.9~$\pm$~ 9.3  &  -4.2~$\pm$~6.0  & -253.6~$\pm$~11.1   &  +6.0~$\pm$~6.3  & 0.483~$\pm$~0.065 & 1.009~$\pm$~0.045  \\
 19--21           &  207    & 20.1                & 105.7~$\pm$~5.3   &  152.1~$\pm$~ 9.7  &  -2.4~$\pm$~6.1  & -245.2~$\pm$~13.4   &  +0.1~$\pm$~6.3  & 0.465~$\pm$~0.084 & 1.020~$\pm$~0.058  \\
 21--23           &  201    & 22.1                & 102.2~$\pm$~5.3   &  148.9~$\pm$~ 9.6  &  +2.4~$\pm$~6.1  & -230.5~$\pm$~13.2   & -10.8~$\pm$~6.3  & 0.438~$\pm$~0.089 & 1.117~$\pm$~0.066  \\
 23--27           &  271    & 24.9                & 100.7~$\pm$~4.5   &  145.0~$\pm$~ 8.1  &  -5.0~$\pm$~5.2  & -251.3~$\pm$~11.4   & -19.0~$\pm$~5.3  & 0.463~$\pm$~0.072 & 1.024~$\pm$~0.050  \\
 27--35           &  263    & 30.5                & 101.8~$\pm$~4.6   &  150.9~$\pm$~ 9.8  & -11.5~$\pm$~5.7  & -235.4~$\pm$~12.9   & -14.9~$\pm$~6.0  & 0.401~$\pm$~0.089 & 1.074~$\pm$~0.064  \\
 35--60           &  301    & 44.5                &  92.4~$\pm$~3.9   &  144.2~$\pm$~ 9.7  & -36.1~$\pm$~6.1  & -229.5~$\pm$~13.4   &  -7.4~$\pm$~6.0  & 0.282~$\pm$~0.130 & 1.115~$\pm$~0.074  \\
\hline
\end{tabular}}
\end{table*}

\begin{center}
\begin{figure*}
\includegraphics[width=\linewidth]{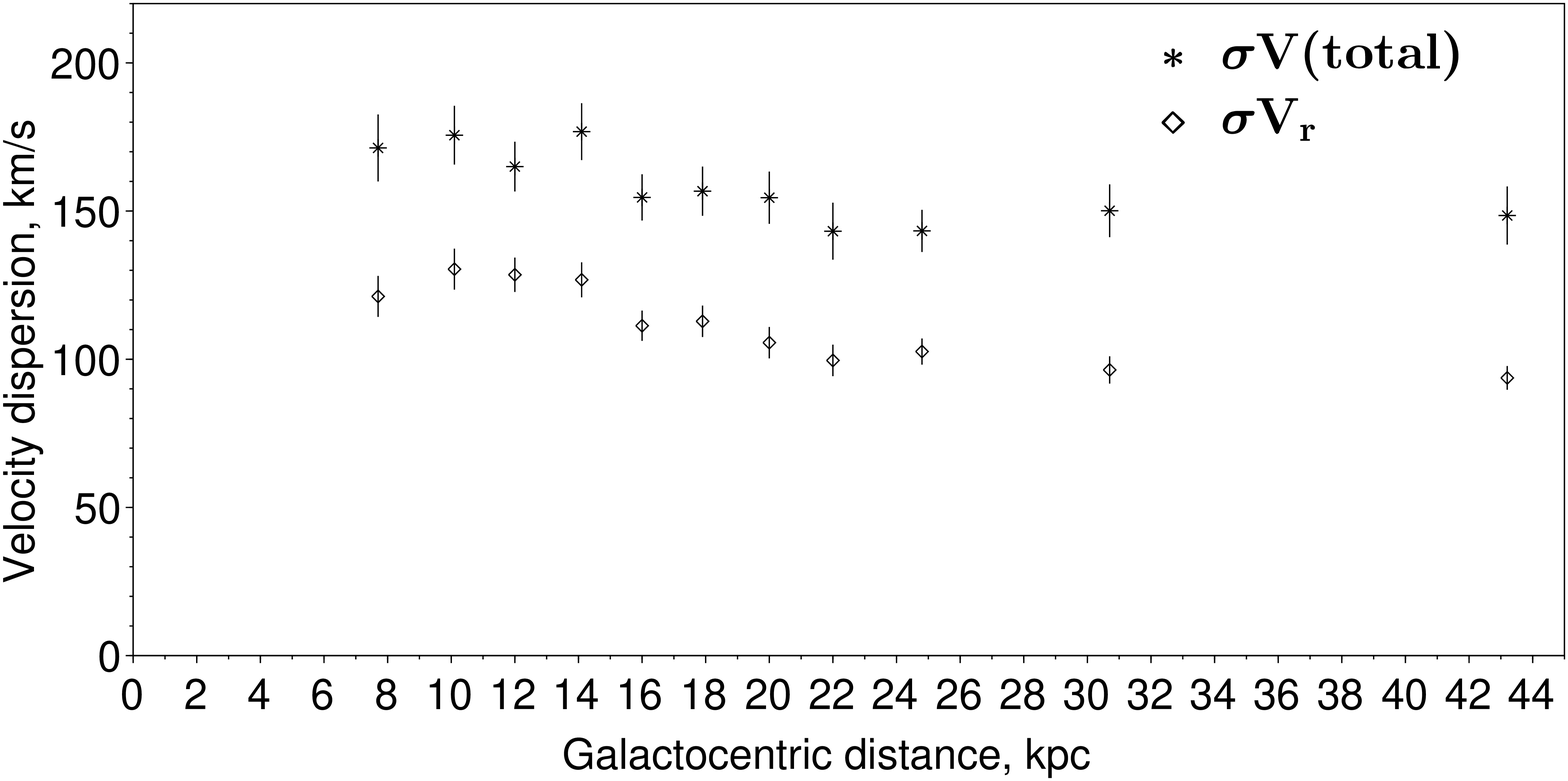}
\caption{Top panel: dependence of the radial ($\sigma_r$, shown with diamond signs) and total ($\sigma V$, shown with asterisks) 
velocity dispersions on Galactocentric distance. }
\label{vdisp}
\end{figure*}
\end{center}

\begin{center}
\begin{figure*}
\includegraphics[width=\linewidth]{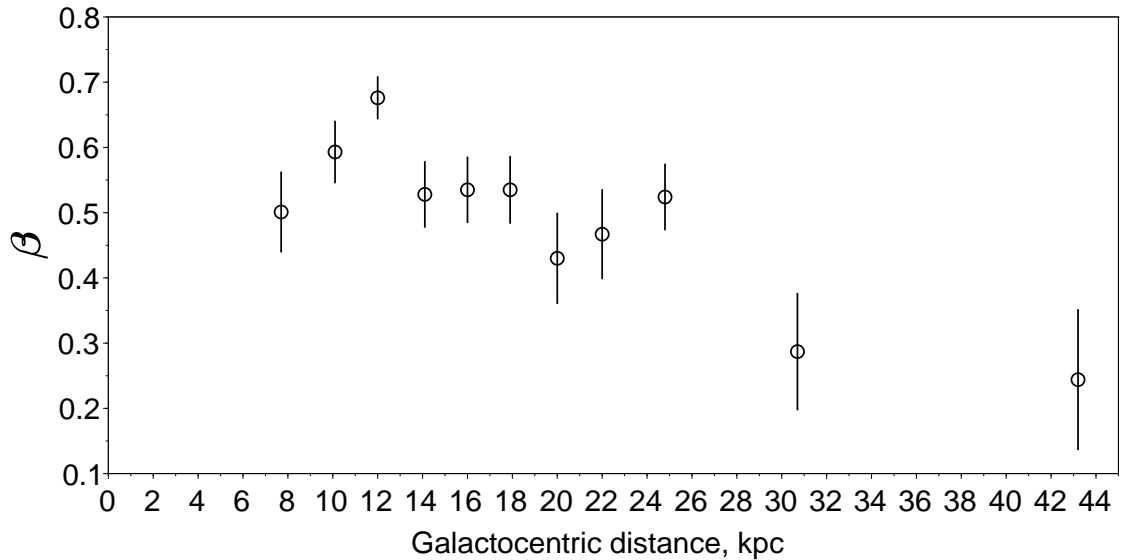}
\caption{Dependence of the anisotropy parameter $\beta$ on Galactocentric distance. }
\label{beta}
\end{figure*}
\end{center}

\begin{center}
\begin{figure*}
\includegraphics[width=\linewidth]{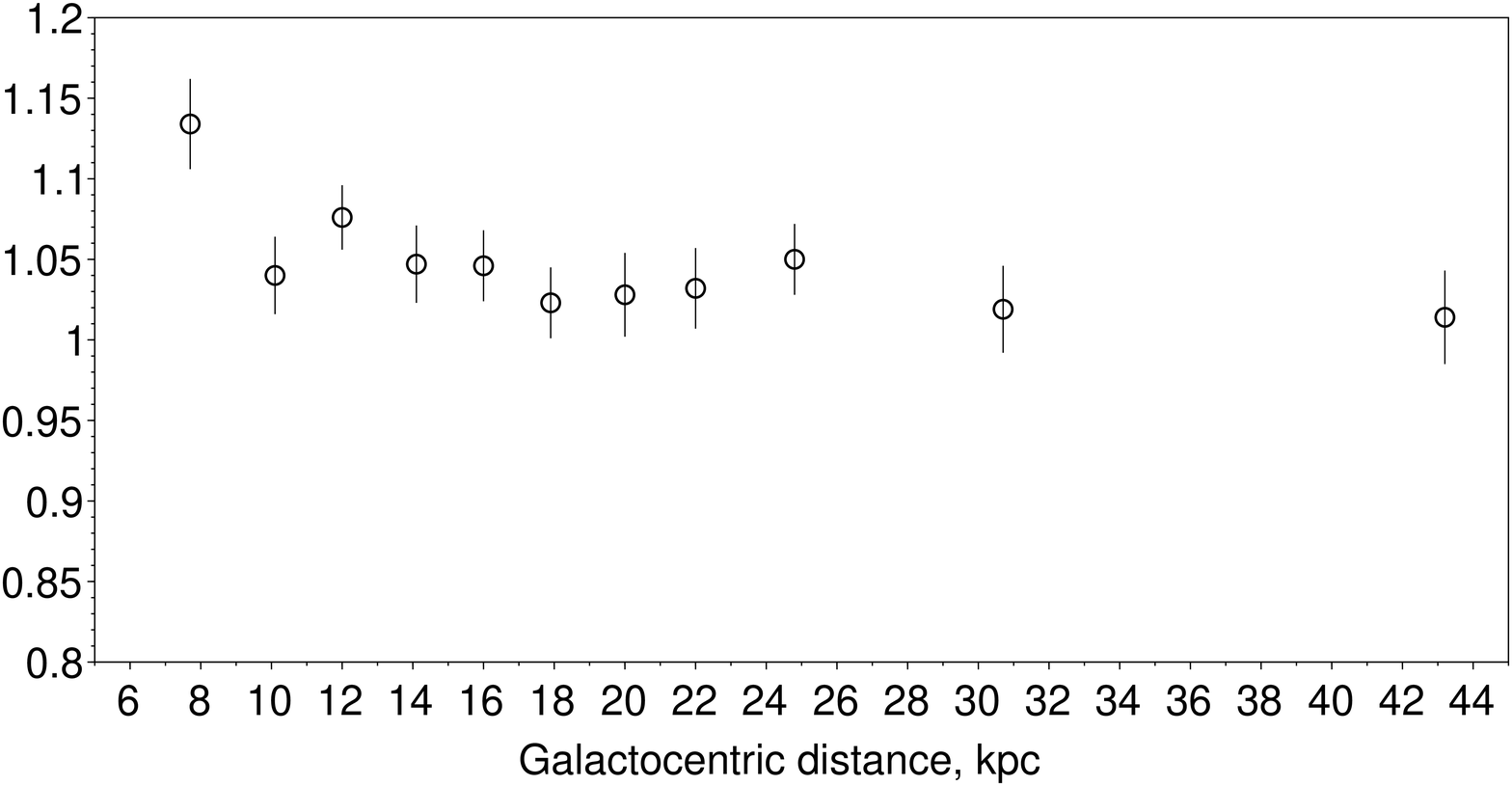}
\caption{Dependence of the distance-scale correction factor $P$ on Galactocentric distance.
The dots and asterisks correspond to
the solutions obtained with $V_0$ treated as
a free and fixed parameter, respectively ($U_0$ and $W_0$ fixed at their values from Table~\ref{all_xue}).}
\label{pfactor}
\end{figure*}
\end{center}

Fig.~\ref{vdisp} shows the dependence of the radial ($\sigma_r$) and total ($\sigma V$) velocity dispersions 
on Galactocentric distance and Fig.~\ref{beta} shows the dependence of
the anisotropy parameter $\beta$ on Galactocentric distance. Fig.~\ref{pfactor} shows the corresponding dependence
for the inferred distance-scale correction factor $P$.

\section{Comparison with the kinematics of RR Lyrae type variables}

\begin{center}
\begin{figure*}
\includegraphics[width=\linewidth]{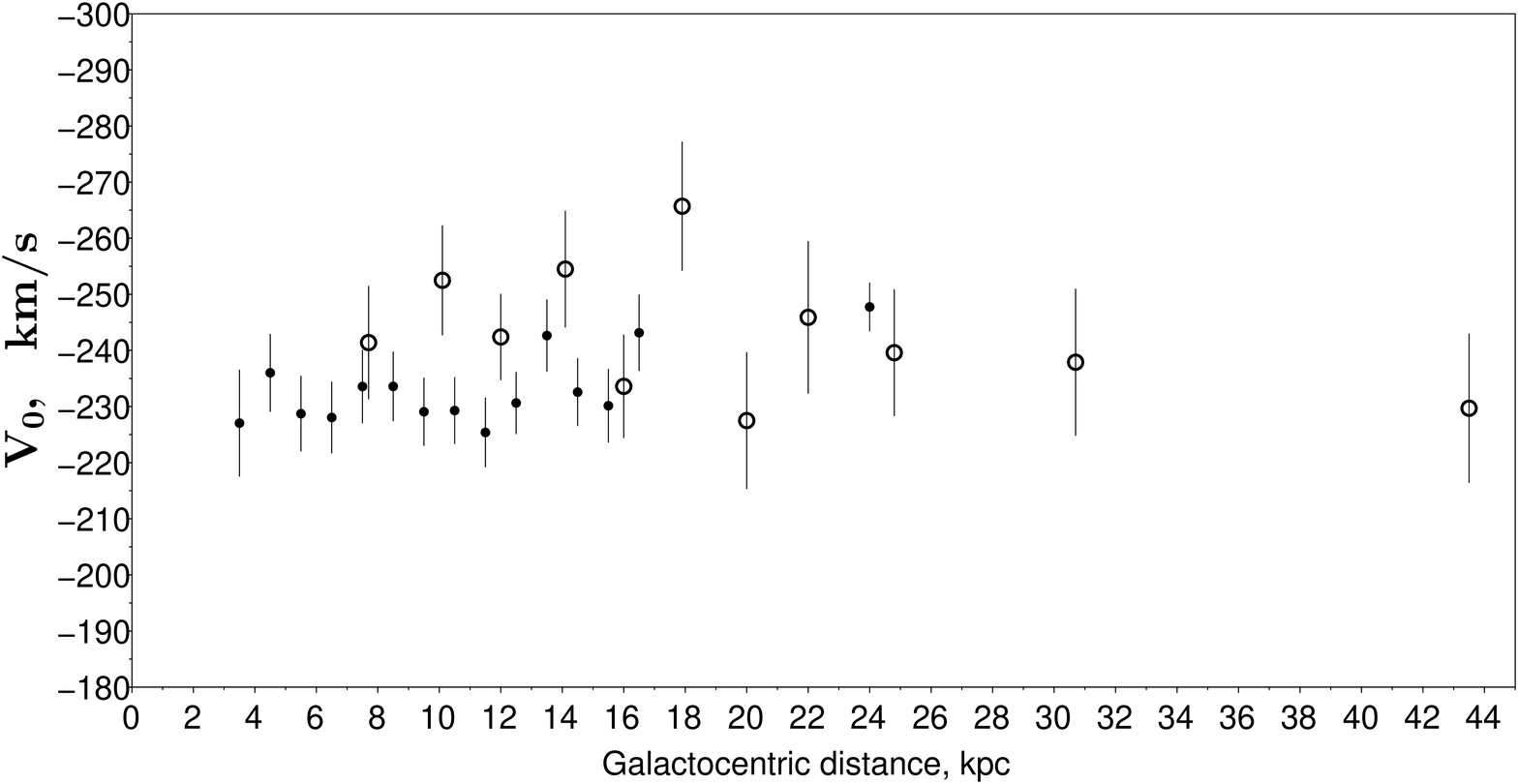}
\caption{Dependence of the mean heliocentric velocity component $V_0$ of BHB stars (the open circles) and RR Lyrae type
variables (the filled circles) on Galactocentric distance.}
\label{v0}
\end{figure*}
\end{center}

\begin{center}
\begin{figure*}
\includegraphics[width=\linewidth]{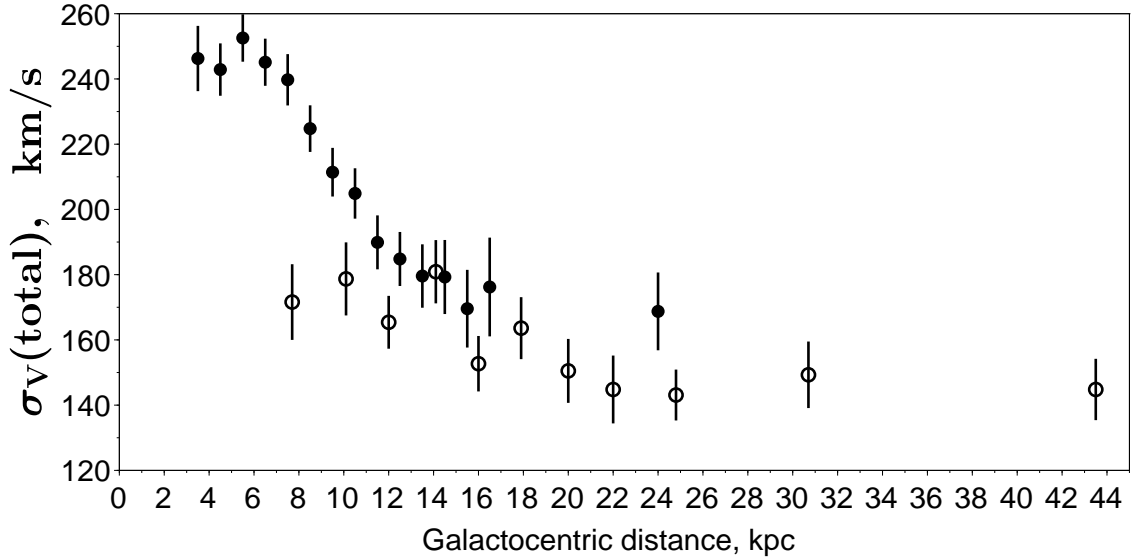}
\caption{Dependence of the total velocity dispersion $\sigma_V$ of BHB stars (the open circles) and RR Lyrae type
variables (the filled circles) on Galactocentric distance.}
\label{sv}
\end{figure*}
\end{center}

Despite their similar evolutionary status, BHB stars and RR Lyrae type variables of the
Galactic halo exhibit somewhat, albeit slightly,
different kinematics. Thus  a comparison of the kinematics of 
these two populations   in the Galactocentric
distance interval from 6 to 18~kpc reveals the following interesting features. 
The average velocity component of BHB stars  in the direction of Galactic rotation,
$V_0$, (typically, $V_0$~=~-241~$\pm$~4~km/s) for BHB stars is slightly smaller in absolute
value than the corresponding velocity for RR Lyrae type variables ($V_0$~=~-222~$\pm$~4~km/s) \citet{Utkin18}.
At the same time, the total velocity dispersion, $\sigma_V$, of BHB stars in the same Galactocentric distance
interval, $\sigma_V$~=~167~$\pm$~4~km/s, is smaller than the total velocity dispersion of RR Lyrae type 
stars in the same Galactocentric distance interval, $\sigma_V$~=~208~$\pm$~4~km/s and the same is true of the
velocity dispersion in the direction of the Galactic centre, $\sigma_V$~=~121~$\pm$~3~km/s for BHB stars
and $\sigma_V$~=~168~$\pm$~5~km/s for RR Lyrae type variables. The anisotropy parameter, $\beta$, also
differs for the two populations: $\beta$~=~0.55$\pm$~0.03 for BHB stars and $\beta$~=~0.71~$\pm$~0.03 for 
RR Lyrae type variables. These are the comparisons of the mean values for the broad interval of Galactocentric 
distances. However, as is evident from Figs~\ref{v0},\ref{sv}, and \ref{beta}, which show the 
Galactocentric distance dependence of $V_0$, $\sigma_V$, 
and $\beta$, respectively, for BHB stars (the open circles) and RR Lyrae type variables (the filled circles),
the differences prove to be rather consistent over the entire range of Galactocentric distances considered.
{ There appears to be no obvious explanation for these discrepancies { and they} are rather surprising given the similar 
evolutionary status of BHB stars and RR Lyrae type
variables, the fact that both of them represent the halo population, and the same technique used to analyse them. The discrepancies
might be due to inevitable contamination of both lists (some variables of other  types may have been erroneously classified as
RR Lyraes and blue stragglers may have infiltrated the BHB list) biasing differently the kinematical behaviour of the two samples.
Another possible cause of the discrepancy may be different degree of the contamination of the two tracer lists by kinematic streams.}

\begin{center}
\begin{figure*}
\includegraphics[width=\linewidth]{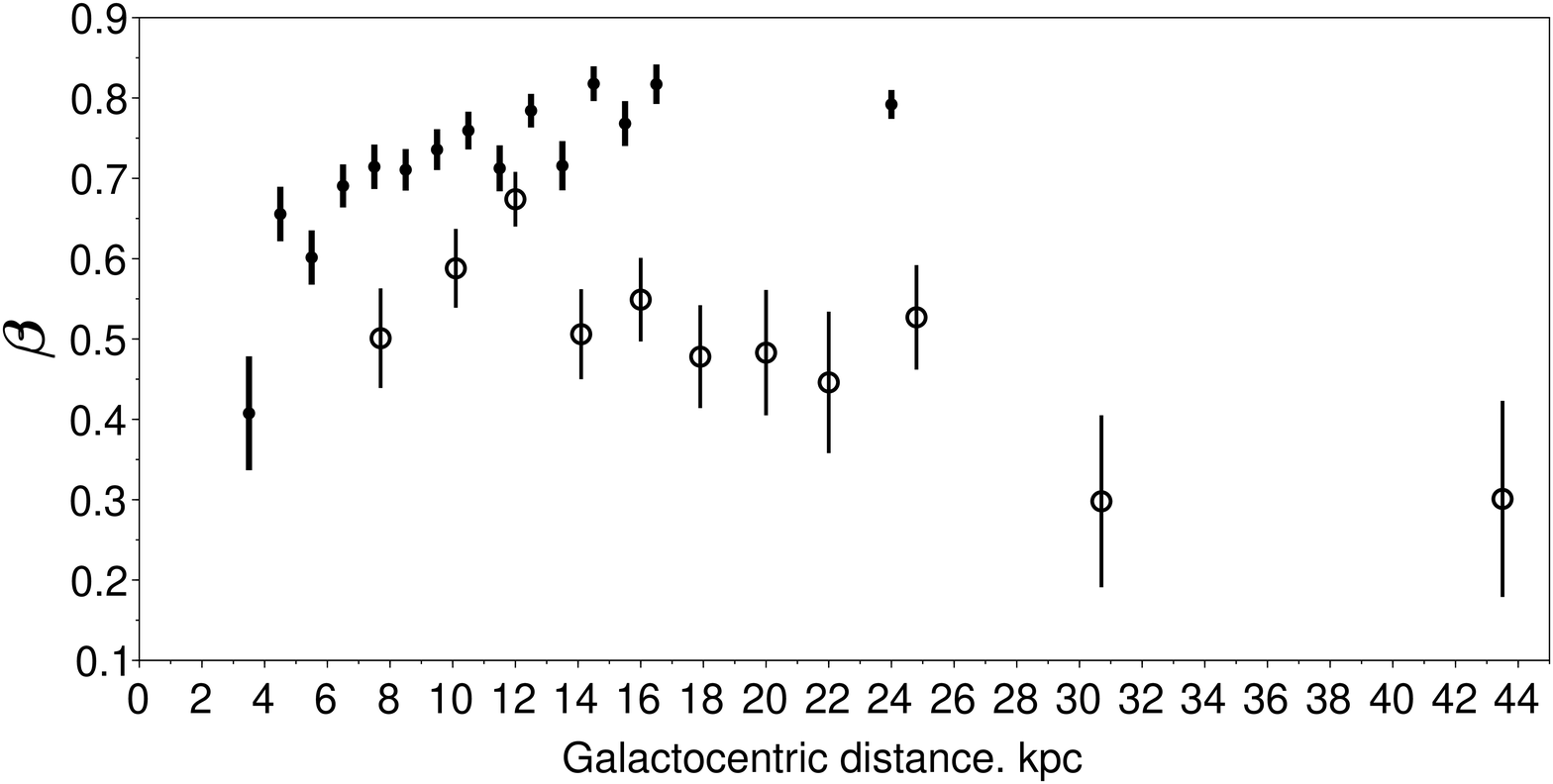}
\caption{Dependence of the anisotropy parameter $\beta$ of BHB stars (the open circles) and RR Lyrae type
variables (the filled circles) on Galactocentric distance.}
\label{beta2}
\end{figure*}
\end{center} 

\section{Conclusions}
\label{sec:conclusion}
We investigated  the kinematics of a clean sample of Galactic halo blue horizontal branch stars with full
6D phase-space data (three space coordinates and three velocity components) using the maximum-likelihood version
of the statistical-parallax technique. The high accuracy of proper motions, radial velocities, and photometric
distance estimates combined with the significantly elongated shape of the velocity dispersion tensor of 
halo BHB stars and the fact that its major axis points toward the Galactic centre allowed us for the first time
to simultaneously determine not only the kinematic parameters of the sample and the photometric distance correction
factor  but also the solar Galactocentric distance $R_0$ from an analysis of the velocity field of halo objects.
We found $R$~=~8.2~$\pm$0.6~kpc, which agrees with other  recent estimates. We also find certain differences
between the kinematics of BHB stars and RR Lyrae type variables in the same Galactocentric distance interval (6 to
18~kpc) despite the similar evolutionary status of the two populations: the velocity ellipsoid of BHB stars is
appreciably less elongated and smaller in size (by about 20~km/s along its major axis) 
than the velocity ellipsoid of RR Lyrae type variables. The results obtained for our BHB sample are
quite robust and stable against mild deviations from Galactocentric spherically symmetric alignment of the velocity ellipsoid.
We find no significant rotation of the sample (rotation velocity does not exceed 2~km/s). Our kinematic
analysis suggests marginal systematic error of SDSS radial velocities of about -5~km/s in the sense that SDSS radial
velocities for our BHB stars are, on the average, underestimated by this amount.

\section{Data availability}

The data underlying this article  were derived from sources in the public domain: VizieR at \url{https://vizier.u-strasbg.fr/viz-bin/VizieR} and SDSS at
\url{www.sdss.org}.

\section*{Acknowledgements}

We are grateful to the anonymous referee for their valuable comments, which helped us to improve the 
paper substantially. This work has made use of data from the European Space Agency (ESA) mission
{\it Gaia} (\url{https://www.cosmos.esa.int/gaia}), processed by the {\it Gaia}
Data Processing and Analysis Consortium (DPAC,
\url{https://www.cosmos.esa.int/web/gaia/dpac/consortium}). Funding for the DPAC
has been provided by national institutions, in particular the institutions
participating in the {\it Gaia} Multilateral Agreement.

This paper made use of SDSS data products. Funding for the SDSS and SDSS-II has been provided by the 
Alfred P. Sloan Foundation, the Participating Institutions, the National Science Foundation, the U.S. Department of Energy, the National Aeronautics and Space 
Administration, the Japanese Monbukagakusho, the Max Planck Society, and the Higher Education Funding Council for England. The SDSS Web Site is 
\url{http://www.sdss.org/}.

The SDSS is managed by the Astrophysical Research Consortium for the Participating Institutions. The Participating Institutions are the American Museum of Natural 
History, Astrophysical Institute Potsdam, University of Basel, University of Cambridge, Case Western Reserve University, University of Chicago, Drexel University, 
Fermilab, the Institute for Advanced Study, the Japan Participation Group, Johns Hopkins University, the Joint Institute for Nuclear Astrophysics, the Kavli 
Institute for Particle Astrophysics and Cosmology, the Korean Scientist Group, the Chinese Academy of Sciences (LAMOST), Los Alamos National Laboratory, the Max-
Planck-Institute for Astronomy (MPIA), the Max-Planck-Institute for Astrophysics (MPA), New Mexico State University, Ohio State University, University of 
Pittsburgh, University of Portsmouth, Princeton University, the United States Naval Observatory, and the University of Washington.

This work is supported by the Russian Foundation for
Basic Research (project nos. 18-02-00890 and 19-02-00611)








\bsp	
\label{lastpage}
\end{document}